\documentclass[aps,prx,twocolumn,longbibliography,eqsecnum,superscriptaddress]{revtex4-2}
\usepackage{titletoc}
\usepackage{amsmath,amsfonts,amssymb,mathtools}
\usepackage{bm}
\usepackage{comment}
\usepackage{graphicx}
\usepackage{sidecap}
\usepackage{verbatim}
\usepackage{hyperref}
\usepackage{xcolor,diagbox}
\usepackage{enumerate}
\usepackage{tikz,cleveref}
\usepackage{bbold}
\usepackage{booktabs}
\usepackage{multirow}
\usepackage{multibib}
\usepackage{adjustbox}
\usepackage{braket}
    \renewcommand{\arraystretch}{2}
\usepackage[english]{babel}
\usepackage[T1]{fontenc}
\usepackage[utf8]{inputenc}
\usepackage{newtxtext}
\usepackage{amsmath}
\usepackage{mathtools}
\usepackage{amsfonts}
\usepackage{bm}
\usepackage{amssymb}

\usepackage{newtxmath}

\usepackage[utf8]{inputenc}
\usepackage[T1]{fontenc}
\usepackage{xcolor}

\usepackage{tabularx}

\usepackage{bm}

\pdfoutput=1
\usepackage{color}
\definecolor{LinkColor}{rgb}{0.156,0.439,0.688}
\usepackage{hyperref}
\hypersetup{
colorlinks=true,
citecolor=LinkColor,
linkcolor=LinkColor,
urlcolor=LinkColor
}
\usepackage{multirow}

\definecolor{todocolor}{rgb}{0.76, 0.0, 0.03}

\usepackage{tikz}
\usepackage{textcomp}

\def\be{\begin{equation}}
\def\ee{\end{equation}}
\def\bea{\begin{eqnarray}}
\def\eea{\end{eqnarray}}

\begin{document}
\title{Beyond the imbalance: site-resolved dynamics probing resonances in many-body localization}
\author{Asmi Haldar}
\email{asmi.haldar@irsamc.ups-tlse.fr}
\affiliation{Univ. Toulouse, CNRS, Laboratoire de Physique Th\'eorique, Toulouse, France}
\author{Thibault Scoquart}
\email{thibault.scoquart@irsamc.ups-tlse.fr}
\affiliation{Univ. Toulouse, CNRS, Laboratoire de Physique Th\'eorique, Toulouse, France}
\author{Fabien Alet}
\email{fabien.alet@cnrs.fr}
\affiliation{Univ. Toulouse, CNRS, Laboratoire de Physique Th\'eorique, Toulouse, France}
\author{Nicolas Laflorencie}
\email{nicolas.laflorencie@cnrs.fr}
\affiliation{Univ. Toulouse, CNRS, Laboratoire de Physique Th\'eorique, Toulouse, France}
\date{\today}
\begin{abstract}
We explore the limitations of using imbalance dynamics as a diagnostic tool for many-body localization (MBL) and show that spatial averaging can mask important microscopic features. Focusing on the strongly disordered regime of the random-field XXZ chain, we use state-of-the-art numerical techniques (Krylov time evolution and full diagonalization) to demonstrate that site-resolved spin autocorrelators reveal a rich and complex dynamical behavior that is obscured by the imbalance observable. By analyzing the time evolution and infinite-time limits of these local probes, we reveal resonant structures and rare local instabilities within the MBL phase. These numerical findings are supported by an analytical, few-site toy model that captures the emergence of a multiple-peak structure in local magnetization histograms, which is a hallmark of local resonances. These few-body local effects provide a more detailed understanding of ergodicity-breaking dynamics, and also allow us to explain the finite-size effects of long-time imbalance, and its sensitivity to the initial conditions in quench protocols. Overall, our experimentally testable predictions highlight the necessity of a refined, site-resolved approach to fully understand the complexities of MBL and its connection to rare-region effects.
\end{abstract}

\maketitle

\newpage
\section{Introduction}
Localization is one of the most salient mechanisms to prevent quantum thermalization. For a single particle in a random potential, Anderson localization~\cite{Anderson_first_PR_1958} is readily probed by considering the exponential-decay of single-particle states. In an equivalent dynamical picture for an Anderson insulator, a wave-packet initially positioned at a certain position in real space does not propagate under time-evolution~\cite{Billy_observation_Nature_2008, Roati_BECond_Nature_2008, Hu_elastic_Nature_2008,lemarie_critical_2010}, showcasing the memory of initial condition for localized systems. For interacting disordered quantum systems, the phenomenon of many-body localization (MBL) can be formally defined through the existence of a quasi-local unitary transformation (that diagonalizes the system) leading to the emergence of local integral of motions (lioms)~\cite{Huse_pheno_PRB_2014, Serbyn_liom_PRL_2013,ros_integrals_2015,Imbrie_2016}, conserved operators with exponentially-decaying support. In an MBL phase, local observables undergoing a quench from an initial state decay to a stationary value which is non-thermal, showing the equivalent memory of initial conditions. The difference with Anderson localization comes from the interactions between lioms whose exponentially-decaying range translates dynamically in a slow logarithmic growth of entanglement and power-law decays of observables (to their long-time values) in the MBL phase~\cite{Bardarson_unbounded_PRL_2012, Vosk_rsrg_PRL_2013, Serbyn_quench_PRB_2014, Lukin_probing_science_2019}. 
This explicit but subtle dynamical distinction between single-particle Anderson localization and MBL is clearly observed in numerical simulations, but is difficult to probe experimentally~\cite{Schreiber_MBL_fermions_2015, Smith_MBL_simulator_2016, Choi_2DMbl_2016, Kohlert_mobility_PRL_2019}, in particular the entanglement entropy~\cite{Lukin_probing_science_2019,chiaro_direct_2022}. 

In practice, localization in many-body systems is often probed through a direct test measurement of the memory of the initial state, namely the imbalance. The idea is to prepare a system in an initial state with a transparent non-equilibrium nature, for instance a real-space density-wave initial state (e.g.  $\ket{m}=\ket{\uparrow \downarrow \uparrow \downarrow \uparrow \downarrow \ldots}$ using without loss of generality a spin representation that will be useful later). The {\it imbalance} with respect to an initial state $\ket m$ can be defined as 
\begin{equation}
    {\cal I}^{m}(t)=\frac{1}{L} \sum_{j=1}^L \bra{m} \sigma^z_j (0) \sigma^z_j (t) \ket{m}
    \label{eq:imb_first}
\end{equation}
where $\sigma^z_j$ is the usual Pauli spin operator,
and the sum runs over all $L$ lattice sites. The quantum dynamics can be encoded in the evolution of the operator $\sigma^z_j(t)=\exp(i {\cal H}_{r} t)\sigma^z_j \exp(-i {\cal H}_{r} t)$. The unitary evolution $\exp(-i {\cal H}_{r} t)$ depends on the Hamiltonian ${\cal H}_r$, which can be a function of the disorder realization $r$.
The imbalance has been considered in almost all experiments probing MBL~\cite{Schreiber_MBL_fermions_2015, bordia_coupling_2016, Smith_MBL_simulator_2016, Choi_2DMbl_2016, luschen_observation_2017, Kohlert_mobility_PRL_2019,yao_observation_2023, li_many-body_2025,hur_stability_2025,lunkin_evidence_2026} and also been widely explored in several computational studies~\cite{Luitz_slow_PRB_2016,torres_generic_2018,doggen_many-body_2018,chanda_time_2020,benini_loschmidt_2021,nandy_dephasing_2021,sierant_challenges_2022,prasad_initial_2022,scoquart_scaling_2025} of MBL, often as the most direct probe of localization, notwithstanding several caveats in its interpretation. For instance, the ultra-slow decay of imbalance (power-law decay $t^{-\beta}$ with very small $\beta$) in numerics~\cite{Luitz_slow_PRB_2016, Agarwal_griffiths_PRL_2015} was interpreted as an indication of eventual thermalization~\cite{doggen_many-body_2018,chanda_time_2020,sierant_challenges_2022} in a disorder regime otherwise thought as MBL from the study of eigenstate properties.

In the formulation Eq.~\eqref{eq:imb_first}, the imbalance explicitly depends on the initial state ($m$), time ($t$) and implicitly on disorder realization ($r$). In order to obtain a simple quantifier (between 0 and 1) of quantum localization, the imbalance is often taken as an average over various distinct sets: long time average (to smooth oscillations), average over $N_s$ initial states, disorder average over $N_r$ different realizations, and finally the real-space average over $L$ sites, which is already performed if one follows the standard definition above.

In the present work, we highlight the importance of carefully distinguishing the different ways of performing these averages to understand the contribution of the various sources of fluctuations to the long-time value of imbalance in the MBL phase. For this, we aim to focus on the site-resolved local autocorrelator as the most elementary object
\begin{equation}
    Z_j^m(t)=\bra{m} \sigma^z_j (0) \sigma^z_j (t) \ket{m},
    \label{eq:Zmintro}
\end{equation} 
and in particular on the distribution of its long-time behavior. We find that the spatial averaging in the imbalance 
\begin{equation}
   {\cal I}^{m}(t)=\frac{1}{L}\sum_j Z_j^m(t)
   \label{eq:Im}
\end{equation}
obscures local resonance effects that may be crucial for explaining various features of the intermediate (sometimes referred to as prethermal~\cite{Bardarson_slow_PRB2019,long_phenomenology_2023,Haldar_slow_PRB2025}) regime between clearly ergodic and localized behavior.  There have been several earlier studies~\cite{Serbyn_quench_PRB_2014, Nandkishore_review_2015, Abanin_review_RMP_2019} on the behavior of local observables similar to $Z_j^m(t)$ in the MBL regime, but mostly focused on either dynamical behavior at short or finite time~\cite{nandy_dephasing_2021,szoldra_tracking_2023,vallejo_single_2025,brighi_probing_2025}, not their long-time limit values, or on expectation values rather than distributions thereof. 

We also find that the local resonance effect we identify is pivotal in the controversial interpretations of dynamical probes of MBL through imbalance. In the XXZ chain defined below, this effect is indeed enhanced by specific initial states, resulting in {\it different finite-size behaviours} for {\it long-time} imbalance. This can result in a misinterpretation of imbalance decay {\it at finite time} --- highlighting further the need of understanding the influence of various ways of averaging. Further, our explicit results on dynamical imprints of local resonances pave the way for a more comprehensive study of dynamical consequences of resonances on all length-scales, which are thought to be key to understand the transition from the MBL to the ergodic regime~\cite{DeRoeck_instability_PRB_2017, Thiery_avalanche_PRL_2018}. Last but not least, these results are relevant for current experiments, and particularly so given the increased precision of site-resolved imagery techniques~\cite{Bakr_singleatom_Nature_2009, Parsons_siteresolved_PRL_2015, Lukin_probing_science_2019}.

To illustrate these important features, we use the standard model of MBL, namely the XXZ spin chain in random field
\be
{\cal{H}}_r=\sum_{j=1}^{L}\left(\sigma^x_j\sigma^x_{j+1}+\sigma^y_j\sigma^y_{j+1}+\Delta \sigma^z_j\sigma^z_{j+1}-2h_j \sigma^z_j\right),
\label{eq:XXZ}
\ee 
where $\sigma^{x,y,z}$ denote the three Pauli matrices, $\Delta$ controls the (Ising) interaction, and the random fields $h_j$ are uniformly drawn in $[-h,h]$. The disorder realization ($r$ in the discussion above) is entirely determined by the set $\{ h_j , j=1\dots L \}$. The factor $2$ in the field-part of ${\cal H}_r$ is chosen to make contact with previous work in spin $1/2$ units. Three important aspects in this Hamiltonian are to be highlighted for the local resonance effect we discuss in the rest of our work. First, the occurrence and precise nature of this effect depend on the form of disorder distribution considered. In particular, the uniform random distribution allows one to have a non-zero probability to neighboring sites to have local fields almost identical (for 2 sites, $h_j \simeq h_{j+1}$, for three sites $h_j \simeq h_{j+1} \simeq h_{j+2}$ etc). Second, the model Eq.~\eqref{eq:XXZ} contains a strong-disorder regime in its phase diagram (see Ref.~\cite{colbois_interaction_2024, colbois_statistics_2024, laflorencie_cat_2025} for recent investigations) where the system shows clear signs of localization using standard estimates, either static, e.g. infinite-time (spectral statistics, eigenstate entanglement, Fock space localization). In this regime, spins close (but not participating) to a local resonance are strongly polarized ('pinned') with a high probability --- which allows to describe analytically the local resonance effect. Finally, ${\cal H}_r$ conserves the total magnetization $M=\sum_j \sigma_j^z$, which will simplify the computations to model analytically this local effect. 

Our work is organized as follows. In Section~\ref{sec:Dynamics}, we first define the relevant dynamical quantities, and use two representative examples to address and illustrate the subtleties of the local magnetization dynamics. We then discuss the central issue of imbalance averaging, presenting various ways of building averages and histograms. This leads us to the observation of a very rich and unexpected multipeak structure in the local magnetization histograms. Section~\ref{sec:toy} presents a simple yet sufficient theoretical model of these numerical observations, based on an analytically tractable toy model of few-body resonances.  
In Section~\ref{sec:consequences} we show how these local resonances can explain significant features of the site-resolved autocorrelators. In particular they allow us to obtain an analytical understanding of the finite-size scaling of the long-time imbalance. Furthermore, we demonstrate that  long-time imbalance critically depends on the chosen initial state. This is highly relevant for ultracold atom experiments, and sheds new light on recent debates triggered by the numerical observation of anomalous slow dynamics, often starting from an initial density-wave (N\'eel) state. 

\begin{figure*}[ht]
    \centering
    \includegraphics[width=2\columnwidth]{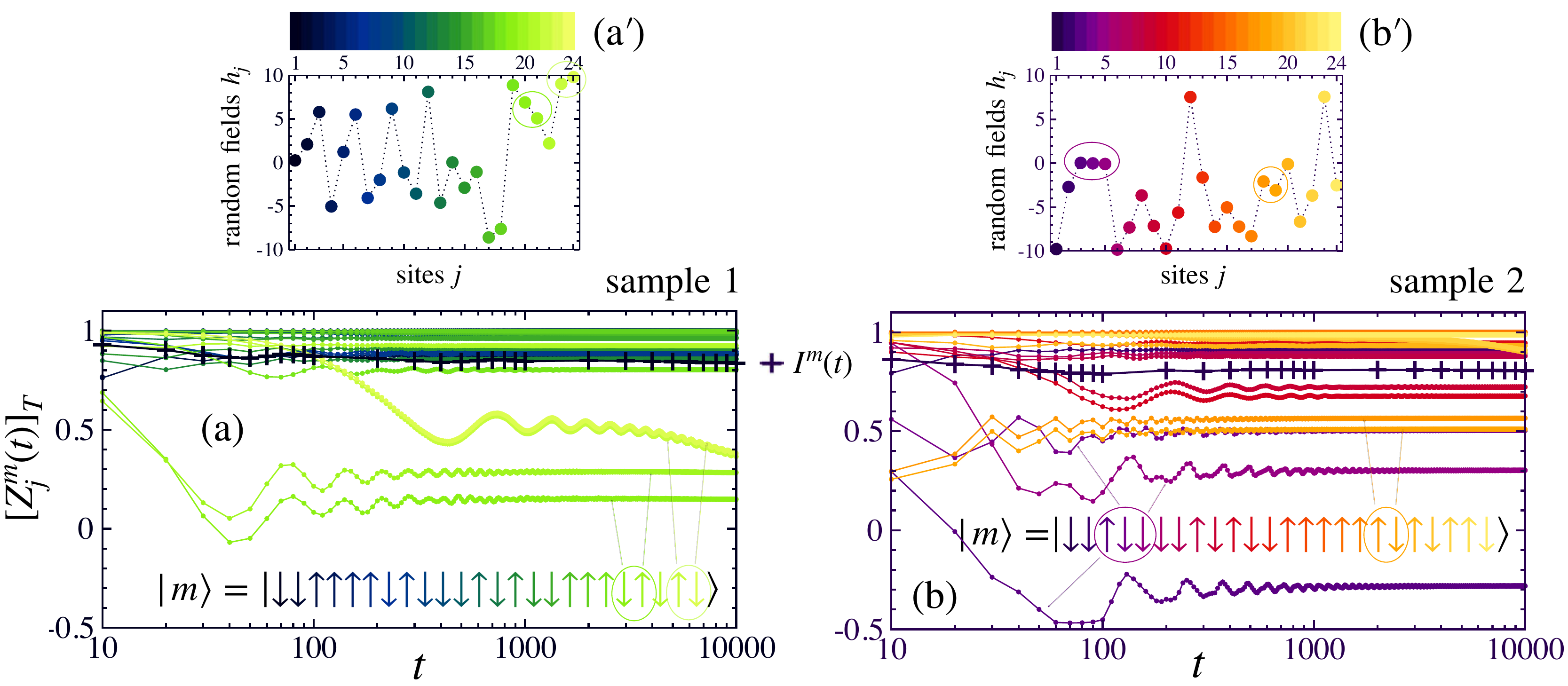}
    \caption{{\it Illustration of what spatial averaging of imbalance hides --- } Local spin dynamics for two typical ($L=24$ sites) samples, at strong disorder ($h=10$) and interaction $\Delta=1$. Starting from an initial product state $\ket{m}$ (shown on each plots), the system follows unitary time evolution Eq.~\eqref{eq:evolution}. The main panels (a-b) show the (time-averaged) site-resolved spin autocorrelators $[Z^{(m)}_j(t)]_T$, see Eq.~\eqref{eq:Zmtime}, for two random field realizations $\{h_j\}$ displayed in panels (a'-b'). The (site and time-averaged) imbalance ${\cal{I}}^{m}(t)$, Eq.~\eqref{eq:Im}, is also shown (+) for comparison. While most sites exhibit an $\mathcal{O}(1)$ plateau at long times---consistent with the site-averaged saturation value of the imbalance in the MBL regime---a few sites (highlighted by colored circles) deviate from this behavior, signaling local resonances that are invisible in $\mathcal{I}^m(t)$.}
    \label{fig:samples}
\end{figure*}

We finish the presentation by providing a visual reading guide of our manuscript. Figs.~\ref{fig:samples} and~\ref{fig:seed9} illustrate through examples what features are hidden by the spatial and initial state averaging of imbalance hides. Figs.~\ref{fig:Histos_L20_LongTimes} and~\ref{fig:Histo_h10} show that the local resonance effect is visible in secondary peaks in some but not all averaged quantities. Figs.~\ref{fig:Histo_2-3br} and~\ref{fig:strongdisorder} illustrate how our toy model captures these secondary peaks, and the final Figs.~\ref{fig:shoulder} and~\ref{fig:FSS_Imb} clearly demonstrate the important consequences of our findings for experiments, regarding state-dependence and the finite-size effects of the imbalance.

Throughout our work, we make sure our results and predictions can be directly tested with currents experiments, by considering realistic system sizes, time windows and number of initial states. Our simulations consist in either full diagonalization of the Hamiltonian \eqref{eq:XXZ} (up to $L=18$ sites) in the zero magnetization sector, or exact time-evolution performed in the Krylov space generated by the initial state $\ket{m}$ (see Appendix~\ref{sec:Krylov}) for chains of sizes up to $L=24$ sites and time up to order $t_{\rm max} \simeq 10^4$). Some of the results required the use of a large number of disorder realizations (up to $N_r=10^5$). 

\section{Imbalance and the various ways of averaging}
\label{sec:Dynamics}

To set up the problem, we consider quench dynamics, starting from product states expressed in the eigenbasis of $\sigma^z$ of the form ${\ket{m}}={\ket{\uparrow\uparrow\downarrow\ldots}}$, and following the unitary time evolution of the state 
\be 
{\ket{\Psi_m(t)}}=\exp(-i{\cal{H}}_r t){\ket{m}}
\label{eq:evolution}.
\ee 
For this form of initial state, the local diagonal autocorrelator $Z^{m}_j(t)$ of Eq.~\eqref{eq:Zmintro} is identically expressed as:
\bea
Z^{m}_j(t)&=&\zeta^m_j\,{\bra{\Psi_m(t)}} \sigma^z_j{\ket {\Psi_m(t)}},
\label{eq:Zm}
\eea
where $\zeta_j^m=\bra m \sigma^z_j\ket m$ encodes the initial local magnetization. 

\subsection{Site-averaging: examples}
\label{sec:site-averaging-example}

We first illustrate how taking a spatial average, as in the imbalance of Eq.~\eqref{eq:Im}, can obscure the rich and nontrivial local features of the autocorrelator $Z^{m}_j(t)$, by presenting the dynamics of two typical samples (specific disorder realization and initial state) at strong disorder ($h=10$) and  interaction strength ($\Delta=1$) for a chain of size $L=24$. Based on static standard estimates  (spectral statistics, eigenstates properties), these parameters place the system on the MBL side of the phase diagram~\cite{pal_many-body_2010,luitz_many-body_2015,sierant_polynomially_2020,colbois_interaction_2024}, despite the clear existence of rare long-range resonances shown in pairwise correlation functions~\cite{laflorencie_cat_2025}, and the apparent ultraslow decay of the imbalance observed when starting from the N\'eel state at this~\cite{sierant_challenges_2022}, or slightly lower~\cite{doggen_many-body_2018}, disorder strength. 

The main panels of Fig.~\ref{fig:samples} display the site-resolved time-averaged autocorrelators (color-coded for different $j$) using the discrete version of Eq.~\eqref{eq:Zmtime}
\be
[Z_j^m(t)]_{T}=\frac{1}{t-t_{\rm min}}\int_{t_{\rm min}}^{t}Z_j^m(\tau){\rm{d}}\tau
\label{eq:Zmtime}
\ee
as it conveniently smooths up short-time irrelevant oscillations, 
together with the site-averaged imbalance Eq.~\eqref{eq:Im} ($+$ symbols). We observe very contrasted behaviors, with a majority of sites showing a very fast convergence to a long-time asymptotic $\sim {\cal{O}}(1)$, in agreement with typical MBL behavior, and a few particular sites which display markedly different features at long time. Their local magnetization fluctuates and decays, retaining little to no memory of its initial value. The resulting imbalance ${\cal{I}}^m(t)$, obtained by site-averaging along the chain, exhibits a standard localized behavior, stabilizing on a long-time plateau reaching a value relatively close to $1$. The dynamics of the site-average imbalance thus completely hides the contrasted behavior of individual, site-resolved autocorrelators. 

The specific random field configurations $\{h_j\}$ (displayed in panel (a)' and (b)') and initial states of these two examples allow us to guess the ingredients needed to explain why some autocorrelators have a wildly different behavior. A closer inspection of the field configurations reveals that these "fluctuating" autocorrelators arise at sites (circled in the insets) characterized by a locally weaker disorder, together with specific initial magnetization patterns. We elaborate in detail on the required conditions through a few-sites toy model in Sec.~\ref{sec:toy}.
\subsection{Initial-state averaging}
\label{sec:state-averaging-example}

Keeping the site resolution, we illustrate now the consequences of another sort of averaging often considered in imbalance measurements: averaging over initial product states, keeping the random field configuration unchanged. This can be achieved either by taking the full trace over the Hilbert space (of size $N_{\rm H}$) or by performing a partial trace over a smaller subset of $N_s \ll N_{\rm H}$ initial states.\\

{\it Full trace --- } Performing the full trace corresponds to evaluating the "infinite temperature" autocorrelator. It can be expanded either in the product-state basis $\{{\ket{m}}\}$, or in the Hamiltonian eigenbasis $\{{\ket{\phi_n}}\}$ (with ${\cal{H}}_r{\ket{\phi_n}}=E_n{\ket{\phi_n}}$):
\bea
{\cal{A}}^{\rm Full}_j(t)&=&\frac{1}{N_{\rm H}}\sum_{m=1}^{N_{\rm H}}Z^{m}_j(t)\nonumber\\
&=&\frac{1}{{ N}_{\rm H}} \sum_{n,n'=1}^{N_{\rm H}} \langle \phi_{n'} | \sigma^z_j | \phi_n \rangle^2 e^{-{\rm i}(E_n - E_{n'})t}.
\label{eq:trace}
\eea
At infinite time, only the terms $n=n'$ are kept, yielding the diagonal ensemble result
\be
{\cal{A}}^{\rm Full}_j(\infty)=\frac{1}{{N}_{\rm H}}\sum_{n=1}^{{N}_{\rm H}}\langle \phi_n|\sigma^z_j|\phi_n\rangle^2,
\label{eq:Aj}
\ee
which is simply an average over all eigenstates. In practice, computing ${\cal{A}}^{\rm Full}_j(\infty)$ requires the entire set of eigenstates (full diagonalization), which is possible for model Eq.~\eqref{eq:XXZ} only up to $L \simeq 18$ sites (for $L=18$, ${{N}_{\rm H}}=48\,620$ in the $M=0$ magnetization sector). 

{\it Partial trace --- } Alternatively, one can also "sample" the sum by restricting to a smaller subset of $N_{s}$ initial states $\ket m$:
\be
{\cal{A}}_{j}^{\rm Partial}(t)=\frac{1}{N_{s}}\sum_{m=1}^{N_{s}}Z^{m}_j(t),
\label{eq:Ajp}
\ee
an approach that is particularly relevant for experiments, which cannot be performed on an exponential number of initial states.\\

    \begin{figure}
    \centering
    \includegraphics[width=.98\columnwidth]{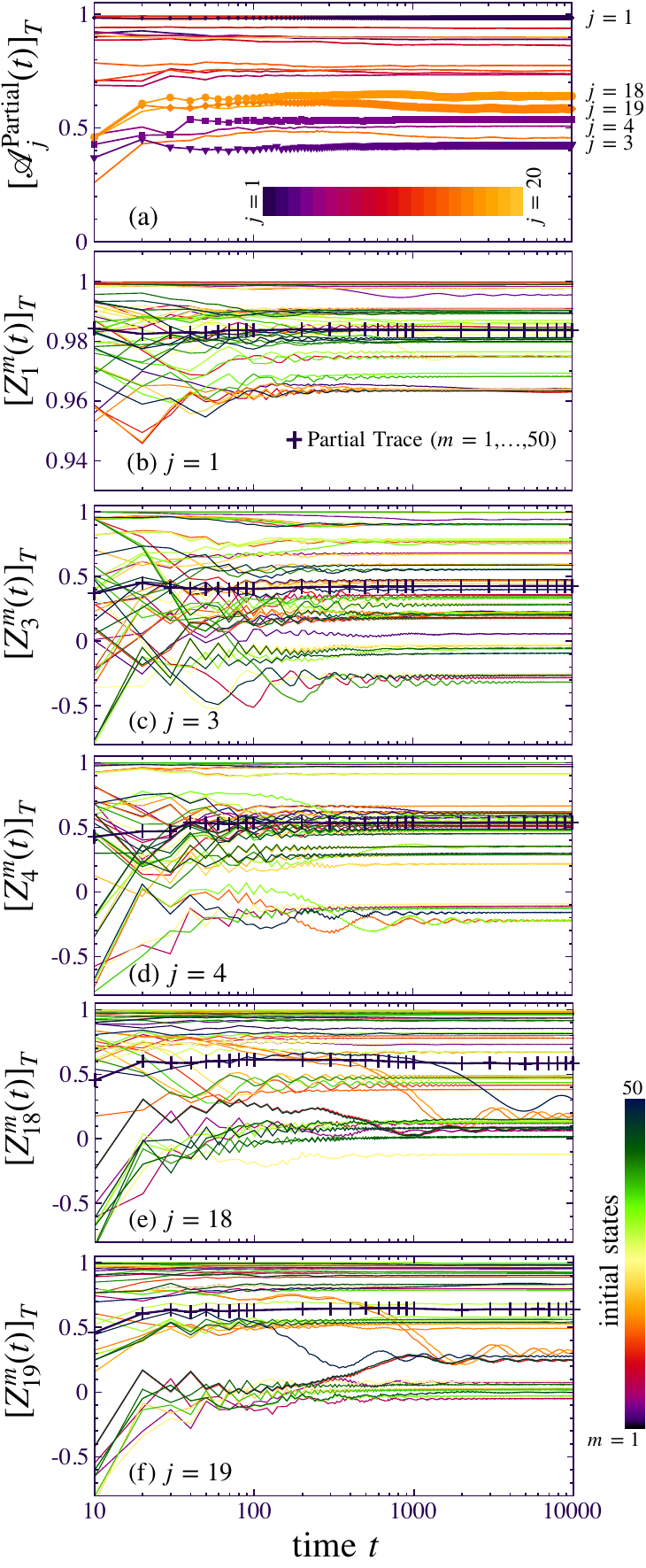}
    \caption{{\it Illustration of what initial-state averaging hides --- } Time-averaged autocorrelators of a single $L=20$ disordered sample having the same first 20 random fields as the one in Fig.~\ref{fig:samples} (b'). 
    Panel (a)  shows the partial traces Eq.~\eqref{eq:Ajp} for all sites $j=1,\ldots, 20$, averaged over ${N_{s}}=50$ initial states. The other panels (b-f) focus each on particular sites ($j=1,\,3,\,4,\,18,\,19$), for which all $[Z^{(m)}_j(t)]_T$ are shown ($m=1,\ldots,\,50$ correspond to different colors).}
        \label{fig:seed9}
\end{figure}

{\it Example of initial-state dependence --- } We illustrate through a specific example the effect of performing an average over initial states (full or partial trace) for a fixed realization of disorder. We consider an $L=20$ sample that shares the same first $20$ random fields as the $L=24$ sample of Fig.~\ref{fig:samples}(b').

The site-resolved partial traces $\mathcal{A}_{j}^{\rm Partial}(t)$ (performed over $N_s = 50$ random initial states chosen to have an energy density close to the middle of the spectrum) are shown in Fig.~\ref{fig:seed9}(a), where a broad spread in the results across the various sites $j = 1,\ldots,20$ is observed. The lower panels (b-f) focus on 5 representative sites ($j=1,\,3,\,4,\,18,\,19$ shown by various symbols in panel a) for which time series for all the 50 initial states are plotted in each panel, together with the partial trace ($+$ symbols). 

Panel (b) illustrates a typical frozen site ($j=1$) which remains almost fully polarized at long time, with a very narrow distribution over different initial states. 
Instead, panels (c-d) and (e-f) correspond to pairs of sites ($j=3-4$) and ($j=18-19$) that have very similar random fields, see Fig.~\ref{fig:samples} (b'), previously identified in Sec.~\ref{sec:site-averaging-example} as fluctuating sites. Interestingly, these sites have a local magnetization dynamics that strongly depend on the initial state. Some initial configurations lead to almost frozen spin polarizations while others show very strong fluctuations. We also identify very similar structures in both pairs of panels (c-d and e-f). We will further elaborate on this strong initial state dependence in the toy model of Sec.~\ref{sec:toy}.

\subsection{Averages and distributions}

The wide fluctuations masked by site or initial state averaging revealed in the examples above strongly suggest focusing on {\it distributions} rather than averages. We begin by formally expressing the different types of averaging that are typically performed when studying imbalance.\\

{\it A question of averaging --- } As most of the studies (numerical or experimental) generally focus on the time evolution of the {\emph{average imbalance}} ${\cal{I}}_{\rm avg}$, it is instructive to express it in the most general manner, as the following triple sum
\be
{\cal{I}}_{\rm avg}(t)=\frac{1}{N_{r}\times N_{s}\times L}\quad\sum_{\mathclap{\substack{r=1,\,N_{r}\\
                              m=1,\,N_{s}\\
                               j=1,\,L}}}
{Z_j^m(t)}
\label{eq:avgI}
\ee
where the average of the local autocorrelators $Z_j^m(t)$ can in principle be performed over three possible sources of fluctuations: 
(1) $N_{r}$  random disordered samples, 
(2) $N_{s}$  initial states, and 
(3) $L$ lattice sites~\footnote{When open boundary conditions are used, it is common to exclude a few sites near the open ends, since those sites are not representative of the bulk physics. In our case, we employ periodic boundary conditions, for which no such restriction is necessary.}.
Interestingly, this triple sum can be expanded in two different ways:
\bea
\label{eq:avg1}
{\cal{I}}_{\rm avg}(t)=\frac{1}{N_{r} N_{s}}\quad\sum_{\mathclap{\substack{r=1,\,N_{r}\\
                              m=1,\,N_{s}
                               }}}
{{\cal{I}}^m(t)}
=\frac{1}{N_{r} L}\quad\sum_{\mathclap{\substack{r=1,\,N_{r}\\
                              j=1,\,L
                               }}}
{{\cal{A}}_j(t)},
\label{eq:avg2}
\eea
where ${\cal{I}}^m(t)$ is the $m-$dependent imbalance Eq.~\eqref{eq:Im}, and ${\cal{A}}_j(t)$ is the site-resolved average autocorrelator, averaged over initial states as defined above in its full (Eq.~\eqref{eq:trace})  or partial (Eq.~\eqref{eq:Ajp}) version. Below we discuss how long-time behaviors of $Z_j^m$, ${\cal I}^m$, and ${\cal A}_j$ are distributed very differently, even though they give the very same average value.\\

\begin{figure}[t!]
    \centering
    \includegraphics[width=\columnwidth]{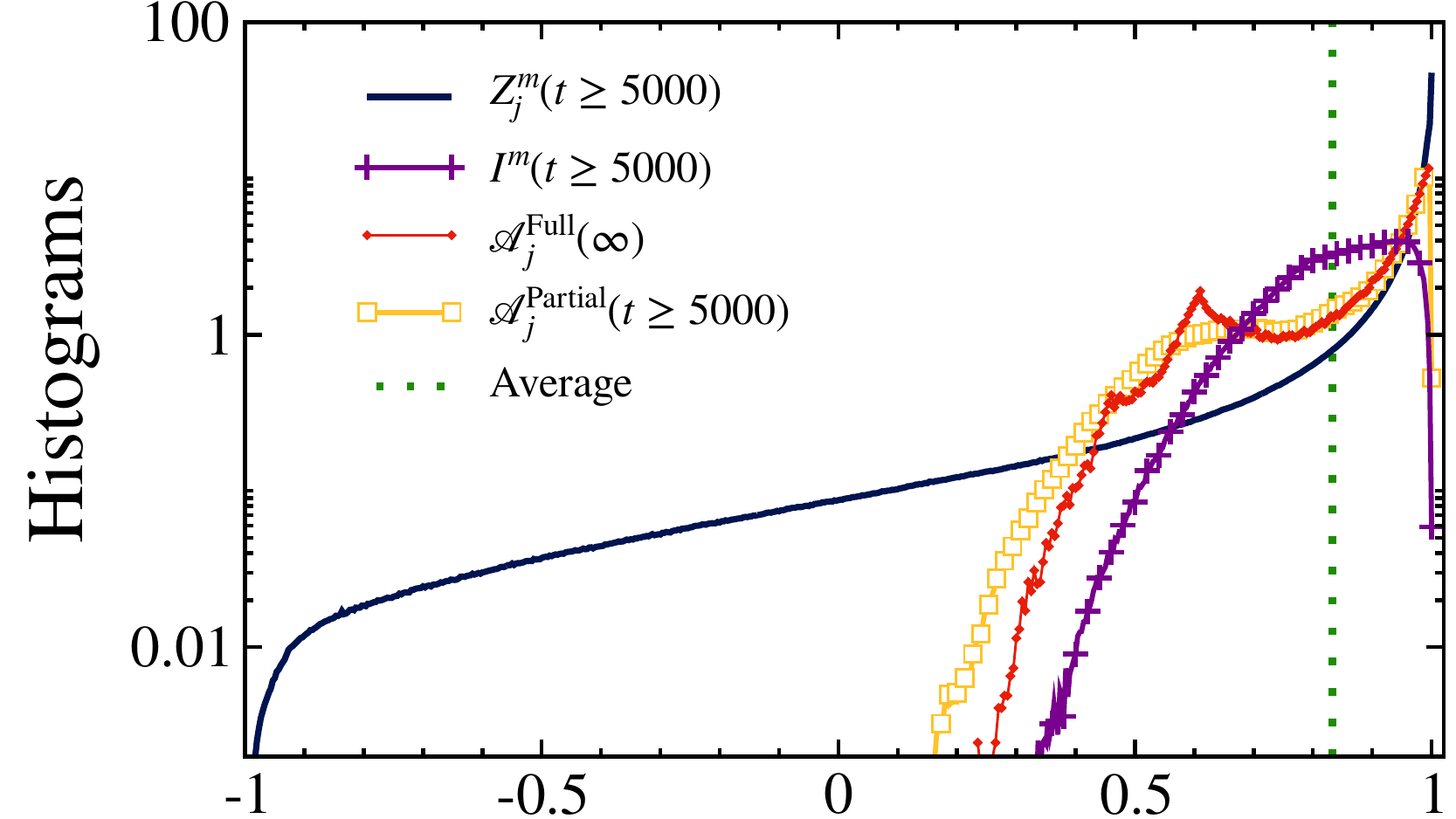}
    \caption{Histograms of the various quantities entering in the average imbalance: Eqs.~(\ref{eq:avgI},\,\ref{eq:avg1},\,\ref{eq:avg2}), all yielding ${\cal{I}}_{\rm avg} \approx 0.83$ (vertical doted line). Long time data ($t \ge 5000$, obtained for $L=20$ sites, $N_{r}=500$ samples, and $N_{s}=50$ initial states) are shown for $h = 10$ and $\Delta = 1$, except for the full trace at $t=\infty$ (red symbols) for which $L=18$ and $N_{r}\approx 10^4$ samples.  $P(Z_j^m)$ (dark thick line) is strongly peaked at 1 with a long tail, indicating a fast suppression of fluctuating sites. $P({\cal{I}}^m)$ (purple crosses) is more peaked close to (but not exactly at) the mean value, shows clear signs of self-averaging. $P({\cal{A}}_j)$ reveals additional structure, notably secondary peaks in ${\cal{A}}_j^{\rm Full}$ (red symbols), also visible but less marked in ${\cal{A}}_j^{\rm Partial}$ (orange squares): these features originate from local resonances, see Sec.~\ref{sec:toy}.} 
    \label{fig:Histos_L20_LongTimes}
\end{figure}

{\it Distributions --- } In Fig.~\ref{fig:Histos_L20_LongTimes}, we present histograms of the various quantities evaluated at long times ($t \ge 5000$) that arise from the different averaging procedures described above. While they all have the same mean, ${\cal{I}}_{\rm avg} \approx 0.83$ for the specific case considered here ($L=18-20$, disorder strength $h = 10$ and Heisenberg interaction $\Delta = 1$), the individual distributions reveal a much richer structure than the average alone, as we describe now.

\paragraph*{(i) $P(Z_j^m)$ --}Strongly peaked at $1$, the distribution $P(Z_j^m)$ then shows a monotonous and fast decay which can be interpreted as follows. For the majority of sites $j$ and initial states $m$, fluctuations are suppressed at long times, and the individual spins remain polarized, essentially retaining the memory of their initial polarization (typical MBL behavior). The rapidly suppressed left tail signals a broad continuum of rare sites that decorrelate at long time from their initial state.

\paragraph*{(ii) $P({\cal{I}}^m$) --}Some self-averaging is expected when the spatial details of $Z_j^m$ are integrated out, see Eq.~\eqref{eq:Im}. Consequently,  the distribution of the individual imbalances is  narrower than $P(Z_j^m)$, but it exhibits a featureless maximum in the vicinity of the mean. The spatial averaging inherent to ${\cal{I}}^m$ has clearly integrated out some relevant microscopic information.

\paragraph*{(iii) $P({\cal{A}}_j$) --}Keeping the spatial resolution but averaging over states is much more interesting. Indeed, the distributions of both full and partial traces exhibit additional richer structures. This is particularly true for ${\cal{A}}_j^{\rm Full}$, which beyond its main peak near 1, develops intriguing secondary maxima. Similar features are also present in ${\cal{A}}_j^{\rm Partial}$ (with a relatively small number of initial configurations, $N_{s} = 50$), although less pronounced (see also Fig.~\ref{fig:shoulder}). These structures are due to local resonances, as we describe and analyze in detail in the next section.

\section{Analytical description of local resonances}
\label{sec:toy}
\subsection{Multi-peak structure: numerical data}
\label{sec:multipeak}
\begin{figure}[t!]
    \centering
    \includegraphics[width=\columnwidth]{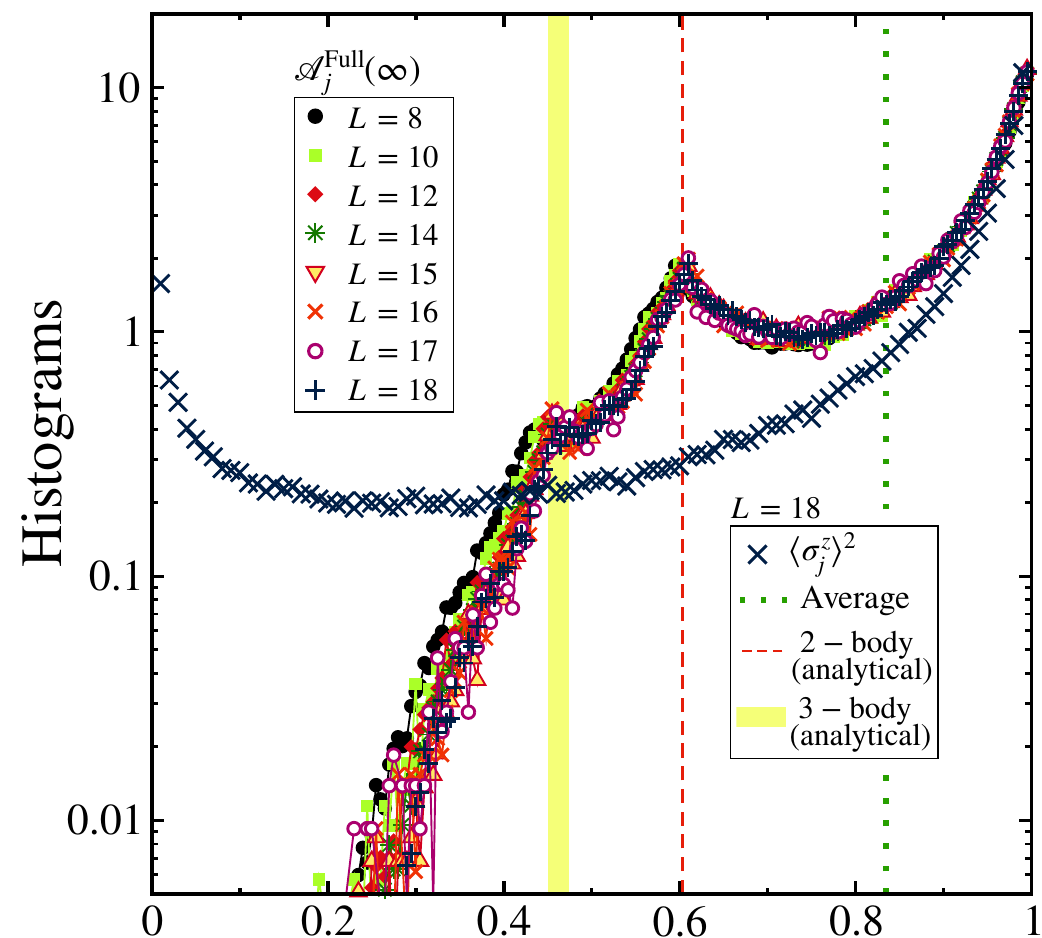}
    \caption{Full diagonalization results for the XXZ Hamiltonian [Eq.~\eqref{eq:XXZ}] at $h = 10$ and $\Delta = 1$.
Histograms of the infinite-time, full-trace observable ${\cal{A}}_j^{\rm Full}(\infty)$ [Eq.~\eqref{eq:Aj}], obtained from $N_r = 10^4$–$10^5$ disorder realizations for various system sizes (as labeled). The distribution of ${\cal{A}}_j^{\rm Full}(\infty)$ differs markedly from that of the squared local magnetizations computed in each individual eigenstate $\langle\phi_n| \sigma_j^z |\phi_n\rangle^2$ (dark crosses, $L = 18$). Spectral averaging reveals a much richer structure (see text for details). The red vertical dotted line signals the 2-body resonance peak, analytically predicted at ${\cal{A}}\approx 0.603$ from Eq.~\eqref{eq:2body_even} (for $\Delta=1$, $L=18$), and the thick yellow line indicates the 3-body resonance at ${\cal{A}}\approx 0.46(1)$ (boundary and middle-site contributions from Eq.~\eqref{eq:3b} cannot be resolved, see also Fig.~\ref{fig:Histo_2-3br}).}
    \label{fig:Histo_h10}
\end{figure}

To study the long-time behavior of local spin dynamics, it is advantageous to focus on the infinite-time limit where the full trace ${\cal{A}}_{j}^{\rm Full}$ takes the simple infinite-temperature average form Eq.~\eqref{eq:Aj}:
\be
{\cal{A}}_j^{\rm Full}(t) \xrightarrow[t \to \infty]{} \frac{1}{N_{\rm H}}\sum_{n}\langle \phi_n| \sigma_j^z|\phi_n\rangle^2
\ee
where the summation is performed (for each disorder sample) over the full set of ${N_{\rm H}}$ eigenstates ${\ket{\phi_n}}$. The distribution of individual contributions from each eigenstate $\langle \phi_n|\sigma^z_j|\phi_n\rangle^2$, displayed in Fig~\ref{fig:Histo_h10}, exhibits a double peak structure with a principal maximum at $1$, and a secondary peak at $0$. 
It is important to note that such distributions of local polarizations, as studied previously for instance in Refs.~\cite{lim_many-body_2016, laflorencie_chain_2020, roy_fock-space_2021}, does not contain much information about the spatial structure of resonances.

Conversely, the distribution of the (full-trace) ${\cal{A}}_j^{\rm Full}(\infty)$ exhibits an even richer set of structures with secondary peaks (beyond the main one at $1$) emerging at specific values, two of which can be clearly distinguished in Fig.~\ref{fig:Histo_h10}. The emergence of these peaks originates from two- and three-body local resonances, which can be captured analytically, as we discuss below.

\subsection{2-body resonances: analytical results}
We find that the secondary peak  occurring at a value $\approx 0.6$ in the distribution of ${\cal{A}}_j^{\rm Full}(\infty)$ in Fig.~\ref{fig:Histo_h10} can be extremely well captured by assuming a local resonance between two neighboring spins. We demonstrate this with 
a toy model that consists of an effective description of 2 sites embedded in a trivially localized product-state background, formally represented by the following "Toy-State" (TS)
\be
{\ket{\rm TS}}={\ket{\uparrow\uparrow\downarrow\ldots}}\otimes{\ket{\circ_1~\circ_2}}\otimes {\ket{\downarrow\uparrow\downarrow\ldots}},
\label{eq:TS}
\ee
where ${\ket{\circ_1~\circ_2}}$ is the pair of (potentially fluctuating) sites for which we want to characterize the quantum dynamics. Assuming that the spins neighboring this pair are frozen and do not fluctuate, we consider the following effective 2-site Hamiltonian for the fluctuating pair:
\be
{\cal{H}}_{\rm eff}=
\sigma^x_1\sigma^x_{2}+\sigma^y_1\sigma^y_{2}+\Delta \sigma^z_1\sigma^z_{2}-2\sum_{j=1}^{2}h_{j}^{\rm eff} \sigma^z_j.
\label{eq:Heff2}
\ee
Here the effective fields $h_{1,2}^{\rm eff}$
account for the effects of external random fields and nearest-neighbor interactions. We provide below the main predictions of the analysis of this effective model, with details of calculations relegated to App.~\ref{sec:AppA}.

Owing to the conservation of total magnetization, the dynamics of the local autocorrelator of the two spins $1,2$ depend on whether their total magnetization 
${\cal{M}}_{1-2}=\frac{1}{2}(\sigma_1^z+\sigma_2^z)$ in the initial state is $0$ or not. For the polarized states ${\ket{\uparrow\uparrow}};{\ket{\downarrow\downarrow}}$ with ${\cal{M}}_{1-2}=\pm 1$, the autocorrelator stays constant $Z^{\pm 1}_{1,2}(t)\approx 1$, $\forall t$. For initial states with vanishing magnetization ${\cal{M}}_{1-2}=0$  $\{ {\ket{\uparrow\downarrow}};{\ket{\downarrow\uparrow}}\}$ on the other hand, the autocorrelators follow an oscillatory behavior:
\be
Z_{1,2}^{0}(t)=1-\frac{2}{1+{\cal{G}}^2}\sin^2(\Omega t).
\label{eq:Zmt}
\ee 
with $\Omega,{\cal G}$ determined by the interaction $\Delta$, the difference of fields $\delta h=|h_1^{\rm eff}-h_2^{\rm eff}|$ and the configurations of the neighboring spins (see expressions in App.~\ref{sec:AppA}). 

At long-time, the time-averaged autocorrelator Eq.~\eqref{eq:Zmtime} for ${\cal{M}}_{1-2}=0$ takes the value 
\be 
\left[Z_{1,2}^{0}(t\gg 1)\right]_{T} \longrightarrow \frac{{\cal{G}}^2}{1+{\cal{G}}^2}.
\label{eq:Ztavg}
\ee
The precise value of ${\cal G}$ 
depends on the configuration of the neighboring spins and are summarized in Tab.~\ref{tab:Zinfty_2body}.


\begin{table}[b!]
  \centering
  \setlength{\tabcolsep}{8pt}    
  \renewcommand{\arraystretch}{1.2} 
  \begin{tabular}{lcc}
    \toprule
    & \multicolumn{2}{c}{Configurations of the neighbours} \\
    \cmidrule(lr){2-3}
    Initial states 
      & $(\uparrow \cdots \uparrow)$ or $(\downarrow \cdots \downarrow)$
      & $(\uparrow \cdots \downarrow)$ or $(\downarrow \cdots \uparrow)$ \\
    \midrule
    $\ket{\uparrow\downarrow}$ or $\ket{\downarrow\uparrow}$
      & $\dfrac{\delta h^2}{1 + \delta h^2}$
      & $\dfrac{(\Delta \pm \delta h)^2}{1 + (\Delta \pm \delta h)^2}$ \\
    $\ket{\uparrow\uparrow}$ or $\ket{\downarrow\downarrow}$
      & $1$ & $1$ \\
    \bottomrule
  \end{tabular}

\caption{Toy model results for the asymptotic values $[Z_{j}^{m}(t\gg 1)]_T$ of the time-averaged autocorrelator in the long-time limit for the two sites $j=1,2$ of the toy model Eq.~\eqref{eq:TS} and Eq.~\eqref{eq:Heff2}. The 16 possible spin configurations, involving the 2 central sites (left column) and the 2 neighbors (top line), are taken into account.}
\label{tab:Zinfty_2body}
\end{table}

\subsection{Average over initial states}
In the 2-site toy model, the full trace required for the computation of the "infinite temperature" autocorrelator Eq.~\eqref{eq:Aj} can be performed analytically. We have
\be
{\cal{A}}_{j}^{\rm Full}(\infty)=\frac{1}{{N}_{\rm H}}\Bigl(\sum_{n=1}^{{{\cal{D}}_0}} \langle \phi_n^{(0)}| \sigma_j^z|\phi_n^{(0)}\rangle^2+\sum_{n=1}^{{{\cal{D}}_p}}\langle \phi_{n}^{(p)}| \sigma_j^z|\phi_{n}^{(p)}\rangle^2\Bigr),
\label{eq:Ajdecomp}
\ee
where the first (second) sum is performed over the subset of states ${\ket{\phi_n^{(0)}}}$ (${\ket{\phi_n^{(p)}}}$) for which the local two-site magnetization is ${\cal{M}}_{1-2}=0$ (${\cal{M}}_{1-2}=\pm 1$).  Combining Eq.~\eqref{eq:Ajdecomp} with the toy model results shown in Tab.~\ref{tab:Zinfty_2body}, the contribution of two-body local resonances is given by:
\bea
\label{eq:A2br}
{\cal{A}}_{\rm 2br}^{\rm Full}(\infty)
&\approx& \frac{{\cal{D}}_{0}}{{2{N}}_{\rm{H}}}\left[\frac{\delta h^2}{1+\delta h^2}+\frac{\Delta^2+\delta h^2+(\Delta^2-\delta h^2)^2}{1+2(\Delta^2+\delta h^2)+(\Delta^2-\delta h^2)^2}\right]\nonumber\\
&+&\frac{{\cal{D}}_{p}}{N_{\rm H}}.
\eea
In the limit $\delta h\to 0$, where the effective 2-site modeling Eq.~\eqref{eq:Heff2} is expected to have the strongest contribution to the secondary peak, we arrive at
\be
\label{eq:A2br_2}
{\cal{A}}_{\rm 2br}^{\rm Full}(\infty)
\approx \frac{1}{N_{\rm H}}\left[\frac{\Delta^2}{2(\Delta^2+1)}{{\cal{D}}_{0}} + {{\cal{D}}_{p}}\right].
\ee
For initial states with minimal total magnetization, the subspace dimensions ${\cal{D}}_0$ and ${\cal{D}}_{p}$  are readily evaluated~\footnote{For even $L$, initial states have $L/2$ spins $\uparrow$ and $L/2$ spins $\downarrow$. If $L$ is odd, we can arbitrarily choose to take $(L+1)/2$ spins $\uparrow$ and $(L-1)/2$ spins $\downarrow$.}. Both tend to $N_{\rm H}/2$, but with different finite-size corrections (see App.~\ref{sec:AppA})
\be
{{\cal{D}}_{0,p}}={N_{\rm H}}\left(\frac{1}{2}\pm \frac{1}{2L-[1+(-1)^L]}\right),
\ee
and an additional dependence on the parity of $L$. Using this result, we find the final finite-size expression for the two-body local resonance contributions to the site-resolved autocorrelator:
\be
{\cal{A}}_{\rm 2br}^{\rm Full}(\infty)\approx \frac{1}{2}+\frac{\Delta^2}{4(\Delta^2+1)}-\frac{1-\frac{\Delta^2}{2(\Delta^2+1)}}{2L-[1+(-1)^L]}.
\label{eq:2body_even}
\ee
For the Heisenberg case $\Delta=1$, one therefore predicts that 2-body resonances should manifest as a peak in $P({\cal{A}}^{\rm Full})$ located at the value $\frac{5}{8}-\frac{3}{8(L-1)}$ (for $L$ even). This analytical prediction is in very good agreement with the numerical results shown in Fig.~\ref{fig:Histo_h10} for $L=18$ (vertical dashed red line at ${\cal{A}}=0.60294$). A more systematic comparison is also provided below in Sec.~\ref{sec:peaks_numanalyt}.

\subsection{Three-body resonances}
One can use a similar approach for three consecutive sites in a region with locally uniform fields. Within 3-site clusters ($j=1-3$), the middle $j=2$ and the boundary $j=1,\,3$ sites display distinct behaviors. The analytical calculation is lengthy but straightforward (for some details see App.~\ref{sec:3br}). The expressions for the full-trace local autocorrelator take the following closed forms:
\begin{equation}
    {\cal{A}}_{\rm 3br}^{\rm Full}(\infty) \approx \begin{cases}
    \cfrac{1}{2}+\Delta^2f_m -\cfrac{g_m}{L}& \text{(middle sites)},\\
    \\
    \cfrac{3}{8}+\Delta^2f_b -\cfrac{g_b}{L}& \text{(boundary sites)},
    \end{cases}
    \label{eq:3b}
\end{equation}
where $f_{m,b}$ and $g_{m,b}$ both depend on the interaction strength $\Delta$, see App.~\ref{sec:3br}.

We report in Tab.~\ref{tab:3b} the expected analytical values for the two- and three-body local resonance peaks, for two representative values of the interaction strength $\Delta=0,\,1$, as a function of system size $L$. For three-body effects, we display both middle and edge (left and right)  spins.


\begin{table}[t!]
  \centering
  \setlength{\tabcolsep}{10pt}      
  \renewcommand{\arraystretch}{1.35}

  \begin{tabular}{l l c c}
    \toprule
    & & $\Delta = 0$ & $\Delta = 1$ \\
    \midrule

    \multicolumn{2}{l}{2-body}
      & $0.5 - \frac{0.5}{L-1}$
      & $0.625 - \frac{0.375}{L-1}$ \\
    \midrule

    \multirow{2}{*}[-0.2ex]{3-body} &
      Middle
      & $0.5 - \frac{0.5}{L-1}$
      & $\approx 0.503 - \frac{0.497}{L-1}$ \\

    & Edges
      & $0.375 - \frac{0.625}{L-1}$
      & $\approx 0.482 - \frac{0.518}{L-1}$ \\
    \bottomrule
  \end{tabular}

  \caption{Expectation values of the single-site autocorrelator at long time
${\cal A}^{\rm Full}(\infty)$ within the toy model descrition of few body resonances: 2-site Eq.~\eqref{eq:2body_even}, and 3-site Eq.~\eqref{eq:3b}, for the non-interacting ($\Delta=0$) and interacting Heisenberg
  ($\Delta=1$) cases.}
  \label{tab:3b}
\end{table}

\subsection{Multi-peak structure: comparison between numerics and toy-model analysis}
\label{sec:peaks_numanalyt}
Building on our more refined analytical understanding of the few-body resonances, we now return to the multi-peak structure  in the site-resolved autocorrelator histograms present in Fig.~\ref{fig:Histo_h10}. We perform a quantitative check of analytical predictions for the peak positions and their finite-size effects for two values of interaction $\Delta$, for large disorder strength $h=10$.

\subsubsection{Heisenberg case $\Delta=1$}
\begin{figure}[b!]
    \centering
    \includegraphics[width=\columnwidth]{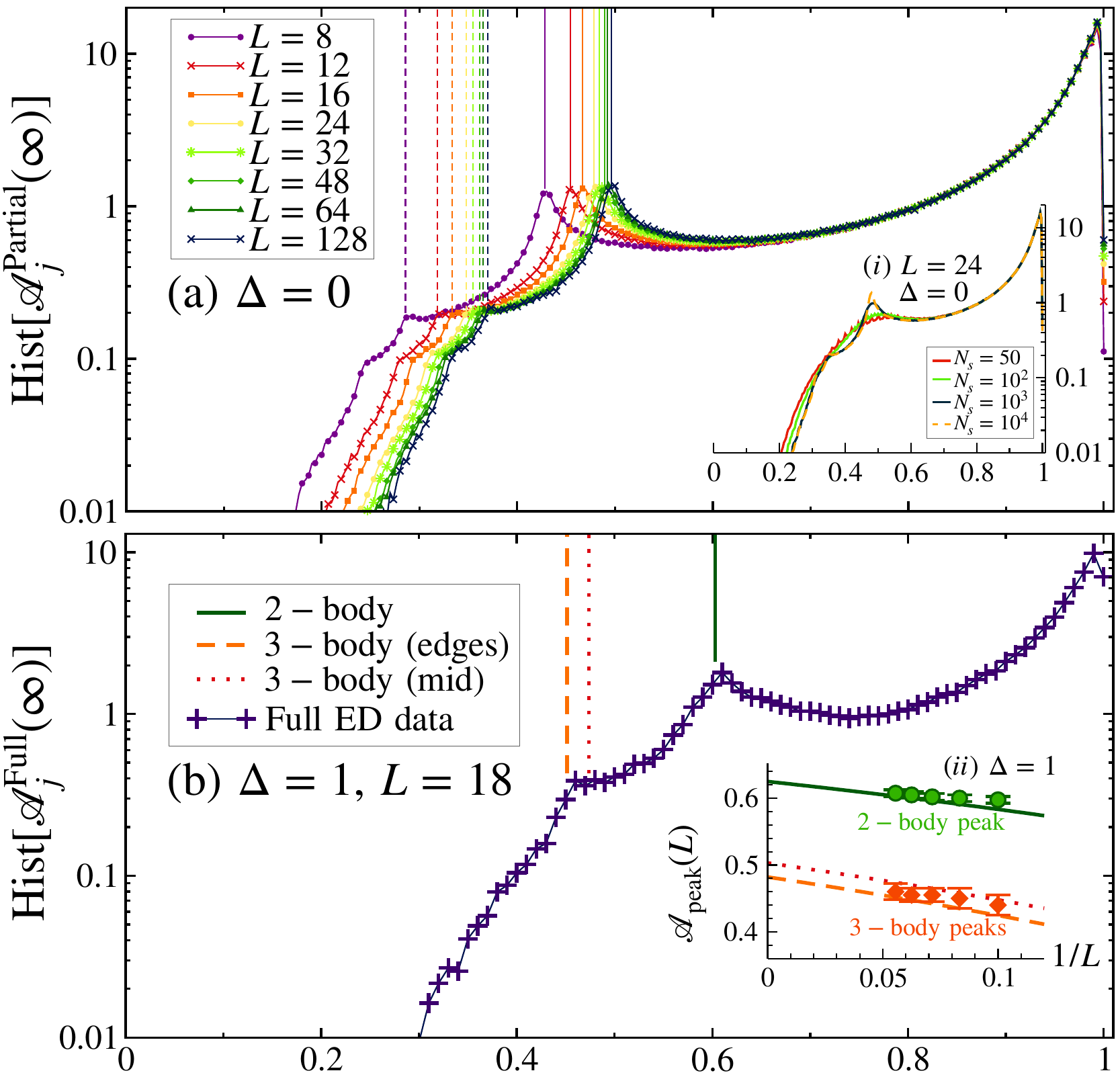}
    \caption{Histograms of the (infinite-time, site-resolved) autocorrelators ${\cal A}_j(\infty)$, averaged over initial states (using full or partial traces), shown for strong disorder $h=10$. (a) Non-interacting case $\Delta=0$: free-fermion results are displayed for system sizes $L=8$–$128$. Data are collected over ${ N}_{r}$ disorder realizations, ranging from $4\times10^4$ to $10^6$ samples. The partial trace is performed over ${N}_{s}=10^4$ randomly drawn basis states $\ket m$. The vertical lines indicate the analytical predictions for 2-body Eq.~\eqref{eq:2body_even} (solid) and 3-body Eq.~\eqref{eq:3b} (dashed) contributions (see also Tab.~\ref{tab:3b}).  Inset $(i)$ demonstrates for $L=24$ how the resonance peaks build in the partial trace upon averaging over an increasing number $N_{ s}$ of initial states. (b) Finite interaction $\Delta=1$ (Heisenberg): same data as in Fig.~\ref{fig:Histo_h10} for $L=18$, together with the comparison with analytical predictions (vertical lines) of Tab.~\ref{tab:3b}. Inset $(ii)$ displays the positions of the few-body resonance peaks {\it{vs.}} $1/L$, again compared to analytical expressions (right column in Tab.~\ref{tab:3b}). The middle {\it{vs.}} edge contributions are hard to resolve numerically.}
    \label{fig:Histo_2-3br}
\end{figure}

We start with the interacting case $\Delta=1$ in Fig.~\ref{fig:Histo_2-3br}(b) where full-trace ED data for $L=18$ sites are shown. The analytical predictions for 2- and 3-body resonances (from Tab.~\ref{tab:3b}) nicely coincide with the positions of the peaks in the numerically obtained histogram. More precisely, the inset (ii) shows how the peaks positions shift with $1/L$, in very good agreement with the 2-body resonance formula Eq.~\eqref{eq:2body_even}. The much less frequent 3-body effect is also very well captured, even though it is very hard to numerically resolve the difference between middle and edge sites, as analytically predicted from  Eq.~\eqref{eq:3b} and Tab.~\ref{tab:3b}.
%

%
\subsubsection{Non-interacting case $\Delta=0$: free-fermion results}
For non-interacting ($\Delta=0$) systems, a wider range of sizes are available ($L=8-128$) thanks to free-fermion calculations. However, performing the full trace can quickly be unattainable as the number of states $N_{\rm H}$ becomes too large (in practice it starts to be numerically costly for $L\gtrsim 20$). We first characterize if it is enough to perform the partial trace, Eq.~\eqref{eq:Ajp},  over a limited number $N_{s}\ll N_{\rm H}$ of initial states. This is shown for $L=24$ in the inset (i) of Fig.~\ref{fig:Histo_2-3br} (a) where we analyze how $P[{\cal{A}}_{j}^{\rm Partial}(\infty)]$ evolves with the number of initial states $N_{s}$. A clear secondary broad maximum is already present for $N_{s}=50$ (similar to $\Delta=1$, Fig.~\ref{fig:Histos_L20_LongTimes}), while the genuine secondary peaks signaling 2-body and 3-body resonances start to be finely resolved for $N_{ s}\ge 10^3$ initial states. We will return to the influence of the number of initial states in the partial sum in Sec.~\ref{sec:finite-sample}.

We are now in position to address precisely finite-size effects on $P[{\cal{A}}_{j}^{\rm Partial}(\infty)]$ for $\Delta=0$. The main panel of Fig.~\ref{fig:Histo_2-3br}(a) shows numerical results for partial traces obtained with  $N_{s}=10^4$ initial states. The histograms are built over a very large number of autocorrelators, typically considering $L\times N_{r}=5\cdot 10^6 -2\cdot 10^7$, with $N_{r}$ the number of disorder realisations. The absence of interaction allows to cover a broad range of system sizes, such that we can perfectly follow the shifts with $L$ of the peaks. Vertical lines denote the analytical expectations for 2-body (full) and 3-body (dashed) effects: the agreement is clearly excellent.
\section{Consequences of the few-site resonances on numerics and experiments}
\label{sec:consequences}

We have illustrated in the previous sections what are few-body local resonances and how they are visible in the histogram of full (or partial) traces of the autocorrelator ${\cal A}_j^{\rm Full/Partial}$.
Building on these results, we address in the present section their theoretical and practical consequences, including in the realistic situations (in experiments and numerics) where MBL is generally studied. This also allows us to discuss the conditions under which these effects are sizeable and can be probed or described by our toy-model analysis. 

We first discuss the statistics of occurrences of these resonances for the box-distribution of random fields in the model Eq.~\eqref{eq:XXZ} at large disorder in Sec.~\ref{sec:stats}. We then address in Sec.~\ref{sec:finite-sample-finite-time} the experimentally-relevant issues of finite-sampling (performing only a partial trace) and finite-time, i.e. how many initial states and which time-window need to be considered for few-body resonances to be detectable experimentally or numerically. We continue in Sec.~\ref{sec:consequence-imb} with a major consequence of the few-body resonances on the {\it finite-size scaling} of imbalance in the infinite- (long-) time limit, highlighting in particular the role of the initial state.

\subsection{Statistics of short-range resonances at large disorder}
\label{sec:stats}

A natural feature of MBL eigenstates is that for strong enough disorder ($h\gg 1$ in Eq.~\ref{eq:XXZ}), most of the spins are strongly pinned, i.e. have a close to perfect polarization $|\bra{\phi_n} \sigma_j^z \ket{\phi_n}| \approx 1$. However, this is not the case for all the sites, as previously seen in Fig.~\ref{fig:Histo_h10}. The toy-model of Sec.~\ref{sec:toy} ascribes this to few-body resonances. What is the probability of such events to occur? Two-body short-range resonances occur when the difference between two consecutive fields $\delta h_j = |h_j-h_{j+1}|$ is small compared to the spin-flip term, which is set to unity in the XXZ Hamiltonian Eq.~\eqref{eq:XXZ}. The probability of $\delta h \approx 0$ depends on the disorder distribution considered. For the random-box distribution $[-h,h]$ used in this work, we have $P(\delta h=0)=1/(2h)$ and thus $P(\delta h \approx 0 )$ scales as $1/h$. Likewise, three-body short-range resonances happen with probability $\sim 1/h^2$, and we thus expect $\ell$-body resonances to be strongly suppressed as $\sim h^{-\ell}$. 

\begin{figure}[t!]
        \centering
    \includegraphics[width=\columnwidth]{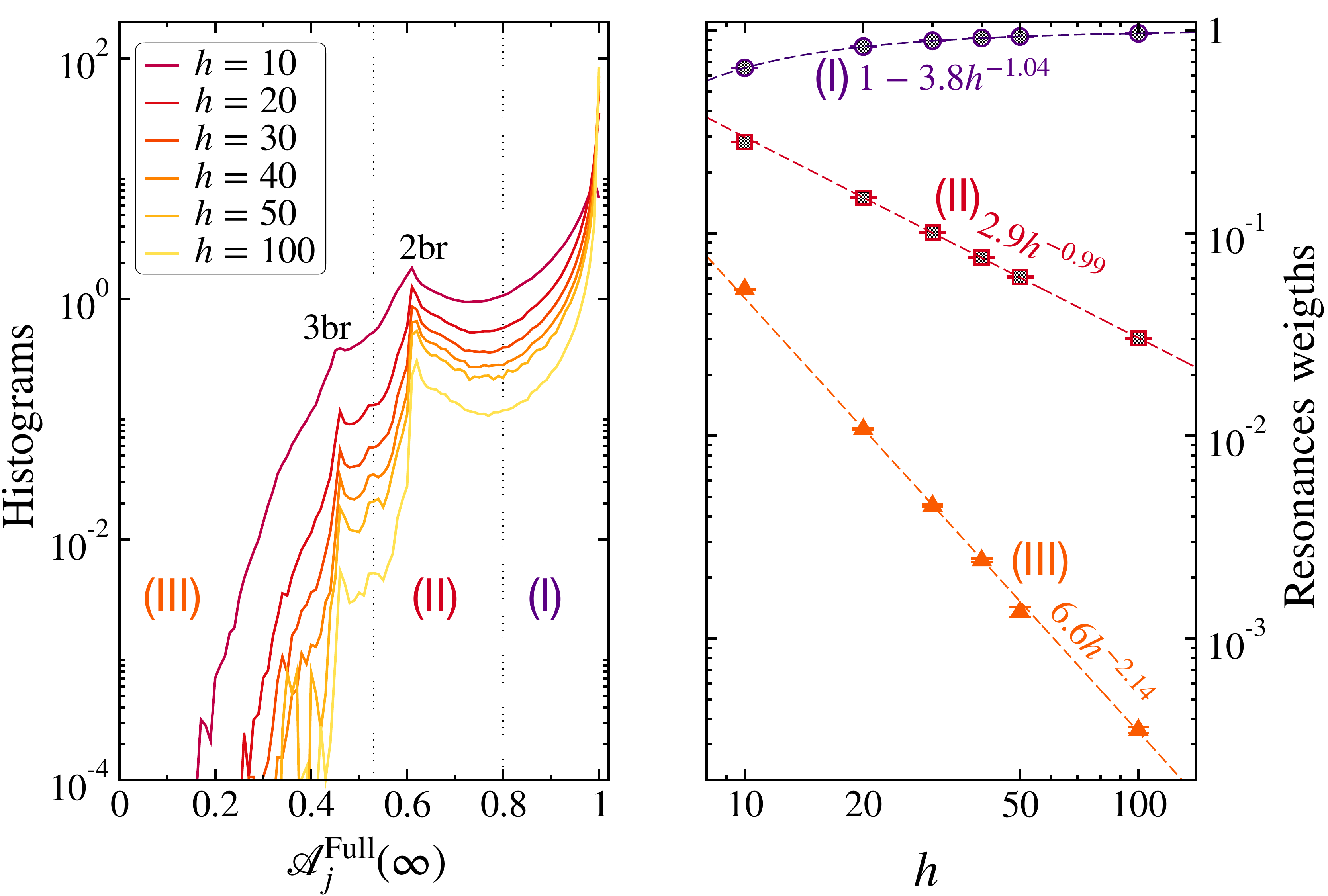}
    \caption{
{\it{Justification of the toy-model description.}}
{Left:} Histograms of the full traces ${\cal A}_j^{\rm Full}$ computed using full ED at very strong disorder ($h\in[10,100]$), for system size $L=14$, averaged over $N_r\sim10^5$ disorder realizations. As $h$ increases, the dominant peak at ${\cal A}\simeq1$ grows, while the secondary two-body (2br) and three-body (3br) resonance peaks progressively lose weight. The two vertical dotted lines separate three distinct regimes: (I) strong polarization, ${\cal A}>0.8$; (II) two-body resonances, ${\cal A}\in[0.53,0.8]$; and (III) mainly three-body (but also potentially more) resonances, ${\cal A}<0.53$.
{Right:} Quantitative analysis of the redistribution of weight among the three regimes (I--III). The different colored symbols represent the integrated weights in each region, together with fits shown in the panel.
}
    \label{fig:strongdisorder}
    \end{figure}

We numerically test this prediction in Fig.~\ref{fig:strongdisorder} by computing the weights of the secondary peaks for various (large) disorder strengths, $h \in [10,100]$. From Fig.~\ref{fig:strongdisorder}(a), we first observe that, as expected from Eqs.~\eqref{eq:2body_even} and~\eqref{eq:3b}, the positions of the peaks are independent of $h$. Furthermore, Fig.~\ref{fig:strongdisorder}(b) shows that the secondary and tertiary peak weights scale as $1/h$ and $1/h^2$, respectively, with good accuracy, thereby confirming the validity of our analysis.
Moreover, as will be exploited later through an analytical modeling of the statistics of resonant {\it{vs.}} frozen sites, Fig.~\ref{fig:strongdisorder}(a) shows that the histograms naturally decompose into a small number of distinct contributions arising from: (i) non-resonant frozen sites, which occur with high probability ($p_{\rm froz.}=1-\alpha/h$); (ii) two-body resonances, which are rare at large $h$ ($p_{\rm 2br}=\alpha'/h$); and (iii) three-body resonances (and higher-order processes), which are even rarer ($p_{\rm 3br}=\alpha''/h^2$).

The random box distribution has the property that the most likely value of $\delta h$ is $0$ -- thus enhancing the probability of such short-range resonances -- with immediate consequence that this effect should increase as disorder is decreased. However, one should note that for the toy-model explanation to be valid, the spins near the short-range resonance should be sufficiently polarized. This implies that the probabilities $P(Z_j^m)$ or $P({\cal{A}}_j)$ have to peak around $1$,
which will no longer be the case if $h$ is too small, as ergodic behavior sets in. We can conjecture that for other disorder distributions, there is an interplay between the value of $P(\delta h=0)$ and the window in the phase diagram where MBL eigenstates display strong polarization, for short-range resonances to be visible and impactful in imbalance measurements.

\subsection{Finite-sampling and finite-time effects}
\label{sec:finite-sample-finite-time}

\subsubsection{Finite number of initial states}
\label{sec:finite-sample}
\begin{figure}
        \centering
    \includegraphics[width=\columnwidth]{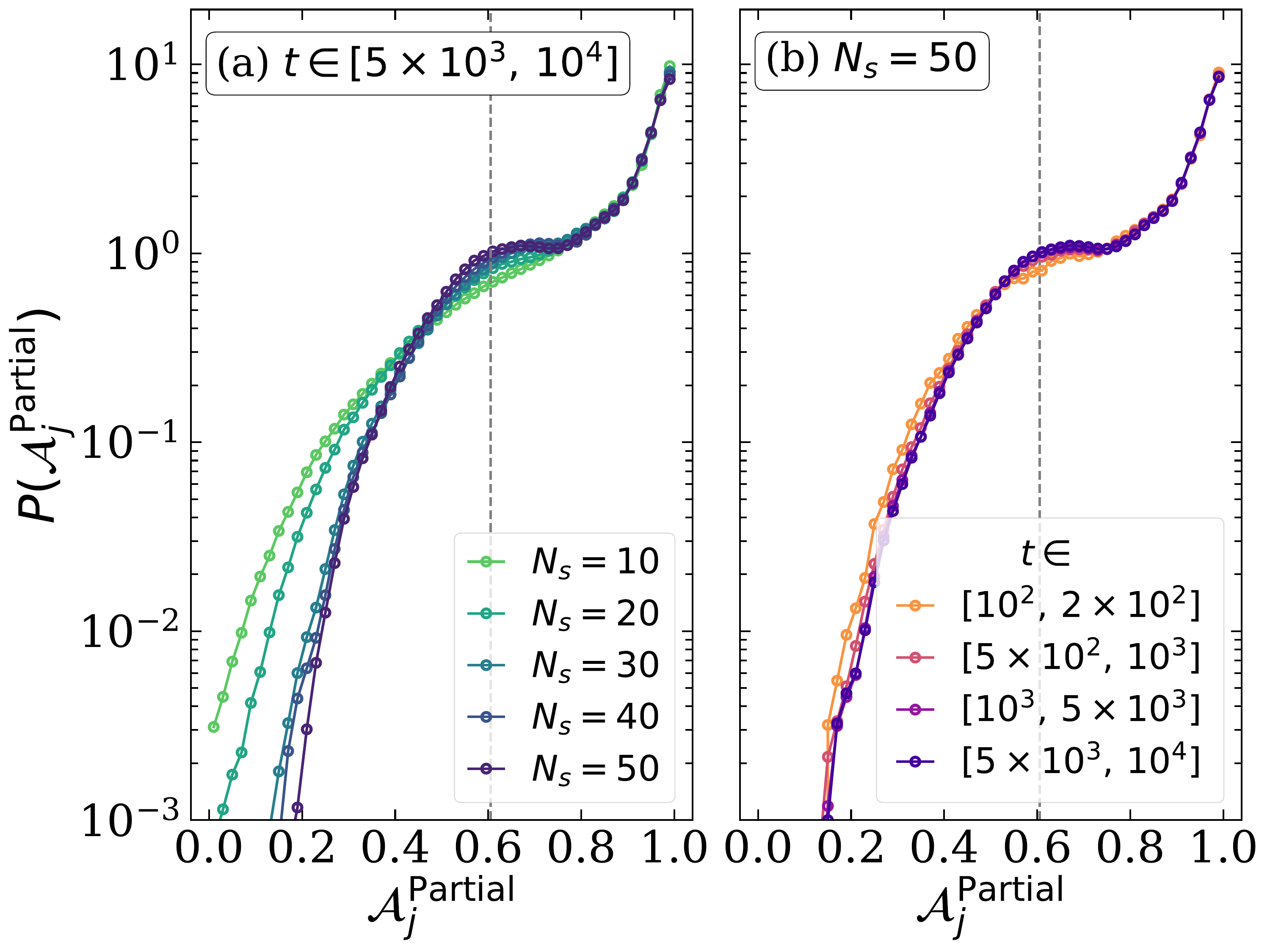}
    \caption{{\it{Finite sampling and finite time effects on the secondary peak building.}} {(a):} Probability distribution of the partial-trace autocorrelator $P({\cal A}_j^{\rm partial})$, where the partial trace is performed over different number $N_s$ of initial states. System size is $L=20$ for $\Delta=1$ and results are obtained from time-averaging for $ 10^4 \ge t \ge 5\times10^3$ for $500$ disorder realizations.
    {(b):} Probability distribution of the partial-trace autocorrelator $P({\cal A}_j^{\rm partial})$ for $N_s=50$ initial states, system size $L=20$ and $\Delta=1$, as a function of various time-windows. The vertical lines indicate the predicted value of the secondary peak from the toy model. Disorder strength is $h=10$.}
    \label{fig:shoulder}
    \end{figure}

Except for very small systems, experimental constraints prevent performing the full trace required to obtain the autocorrelator ${\cal A}_j^{\rm full}$, leaving only the partial estimator ${\cal A}_j^{\rm partial}$ accessible.
 A relevant question is thus: how many initial states (per realization of disorder) are needed so that ${\cal A}_j^{\rm partial}$ captures the main features of ${\cal A}_j^{\rm full}$, included in particular secondary peaks\,? The toy-model analysis provides a clear answer: for a 2-body resonance to be well captured, one needs to have a good sampling of the $2^4=16$ local configurations, or more precisely to the (equally-likely) four contributions of Tab.~\ref{tab:Zinfty_2body}. We thus expect that when $N_s \gg 4$ (for instance, starting from $N_s \sim 50$), the secondary peak should be reasonably well captured. The onset of the secondary peak was already visible in Fig.~\ref{fig:Histo_2-3br} through a shoulder in the distribution $P({\cal A}_j^{\rm partial})$ located at the position of the peak in $P({\cal A}_j^{\rm full})$. We test systematically the role of $N_s$ in the build-up of this shoulder in Fig.~\ref{fig:shoulder}(a), where we see that a structure (shoulder-like for small number of initial states $N_s$, and more rounded as $N_s$ increases) starts to develop around the position of the expected secondary peak of $P({\cal A}_j^{\rm full})$ (denoted by a vertical dashed line in the figure). From these data, we estimate that with $N_s \simeq 20-30$ initial states, this feature can be directly probed experimentally. If we use a similar argument for the 3-body short-range resonance (resulting from $2^5=32$ local configurations), we expect that the tertiary peak could be visible experimentally for a larger number of initial states. However the 3-body short range resonances occur with an overall probability scaling as $1/h^2$, requiring a very large number of disorder realizations to be visible, which explains why it is not detectable in Fig.~\ref{fig:shoulder}(a) where $N_r=500$.

\subsubsection{Finite time in simulations/experiments}
Another important experimental aspect is that the system can stay fully isolated from the environment only for a limited amount of time (ranging up to a few hundreds-thousands of typical hopping/spin-flip times). We thus need to determine at what typical times the influence of the few-body short-range resonances (which we argue produce characteristic effects in infinite-time quantities (such as ${\cal A}_j^{\rm full}$ or ${\cal A}_j^{\rm partial}$ ) can be detected. The toy-model Eq.~\eqref{eq:Zmt} predicts oscillations around a mean value with a frequency $\Omega$ or order $1$ for typical short-range 2-body resonances, thus corresponding to typical set-up times of order $1$ (in natural units). While some oscillations are visible on these time scales in the time-traces of Fig.~\ref{fig:samples} and Fig.~\ref{fig:seed9}, one also observes other dynamical effects affecting the long-time values on much longer time scales (some being of the order of a few hundreds and sometimes thousands time units). Such longer-lived oscillations cannot be captured  by a short-range 2-body resonance model which involves energy differences of order $1$, given by the energy scale of direct coupling present in the bare XXZ Hamiltonian Eq.~\eqref{eq:XXZ}. There are indeed {\it longer-range}  resonances (involving two or more spins) resulting from more complex many-body processes. Explicit examples of such long-range many-body resonances taking the form of cat states were recently provided in Ref.~\cite{laflorencie_cat_2025}, with exponentially small energy differences corresponding to very long-time scales. 

In order to go beyond the toy model, which does not allow to completely predict the complex dynamics of the local autocorrelators at {\it{finite time}}, we perform numerical experiments where we study the buildup of the secondary peak in ${\cal A}^{\rm partial}(t_{\rm max})$ as a function of final time $t_{\rm max}$. This is represented in Fig.~\ref{fig:shoulder}(b) 
using a realistic number of initial states ($N_s=50$).  
In addition to the secondary peak seen to emerge with increasing $N_s$ in Fig.~\ref{fig:shoulder}(a), a similar feature appears as the observation time grows: the rounded secondary peak gradually forms in the distribution of the partial trace $P({\cal A}_j^{\rm partial})$ at the predicted value.
We thus conclude that this feature becomes experimentally detectable on a moderately short time, $t \approx 500$, using a small number of initial states, $N_s \simeq 20\text{--}30$.

\subsection{Consequences for average imbalance and finite-size scaling}
\label{sec:consequence-imb}
When the average imbalance ${\cal I}_{\rm avg}$ is the result of an average over sites, disorder, and {\it{all}} eigenstates (full trace autocorrelator ${\cal A}_j^{\rm Full}$), we expect it to inherit the finite-size drift discussed in Fig.~\ref{fig:Histo_2-3br} and Tab.~\ref{tab:3b}. At long (infinite) time, we thus anticipate that ${\cal I}_{\rm avg}$, when averaged over all initial states, will display similar finite-size corrections as of ${\cal A}^{\rm Full}(\infty)$, given in Tab.~\ref{tab:3b}, i.e. with a {\it negative} $1/L$ correction, meaning that the thermodynamic value ${\cal I}_{\rm avg}(L=\infty)$ should be reached from below. 
This expectation holds as long as the assumptions of the toy model are satisfied, namely that spins neighboring a local resonance do not fluctuate and are not themselves involved in a (possibly longer-range) resonance. For $h = 10$ and $\Delta = 1$, we find that this condition is indeed met, as evidenced by the clear primary peak at $1$ in Fig.~\ref{fig:Histo_h10}.

\subsubsection{Statistical modeling at strong disorder}
Assuming a simplified statistical description of a strongly disordered chain ($h\gg 1$) where spins either belong to a few-site resonant region 
or are almost totally frozen $\langle \sigma^z\rangle \simeq \pm 1$, we expect the average infinite-time imbalance to be given by
\bea
{\cal{I}}_{\rm avg} (L) & \approx & \frac{1}{L}\left(\,\,\sum_{\mathclap{\substack{{\rm frozen}\\
                              {\rm sites}}}} {\cal{O}}(1)\,+\,\sum_{\mathclap{\substack{{\rm fluctuating}\\
                              {\rm sites}}}} 
                               {\overline{Z_j(\infty)}}\,\right),\nonumber\\
                               &\approx& \left(1-\frac{\alpha}{h}\right) \, +\, 
                               {\overline{Z_{\rm fluct.}(\infty)}}\,\frac{\alpha}{h}+{\cal{O}}\left(\frac{1}{h^2}\right),
                               \label{eq:Iavg_toy}
\eea
where $\alpha$ is an ${\cal{O}}(1)$ non-universal number.
Within this strong disorder description, justified by data shown in Fig.~\ref{fig:strongdisorder}, we only consider two types of behaviors: (i) almost perfectly polarized (non-fluctuating) spins, present with high probability ($p_{\rm froz.}=1-\alpha/h$) ; (ii) 2-body resonant sites, which are rare at large $h$ ($p_{\rm fluct.}=\alpha/h$). We neglect 3-body (and beyond) effects which come with much smaller probability.

There are various ways to obtain the average infinite-time autocorrelator of the fluctuating sites ${\overline{Z_{\rm fluct.}(\infty)}}$ in Eq.~\eqref{eq:Iavg_toy}. If one does the full trace, we just need to replace ${\overline{Z_{\rm fluct.}(\infty)}}$ by ${\cal{A}}_{\rm 2br}^{\rm Full}(\infty)$ given by Eq.~\eqref{eq:2body_even}, the averaging over disordered samples will just then brings the above probabilities $p_{\rm froz.}$ and $p_{\rm fluct.}$. However, we can also consider the more realistic options of starting from single specific or random initial states, building on the results of Tab.~\ref{tab:Zinfty_2body}. All our results on the finite-size corrections to the average imbalance are shown in Fig.~\ref{fig:FSS_Imb}, which are also described by simple fitting functions provided in Tab.~\ref{tab:Ifits}.

\subsubsection{Initial state dependence}
\begin{figure}[b!]
        \centering
    \includegraphics[width=\columnwidth]{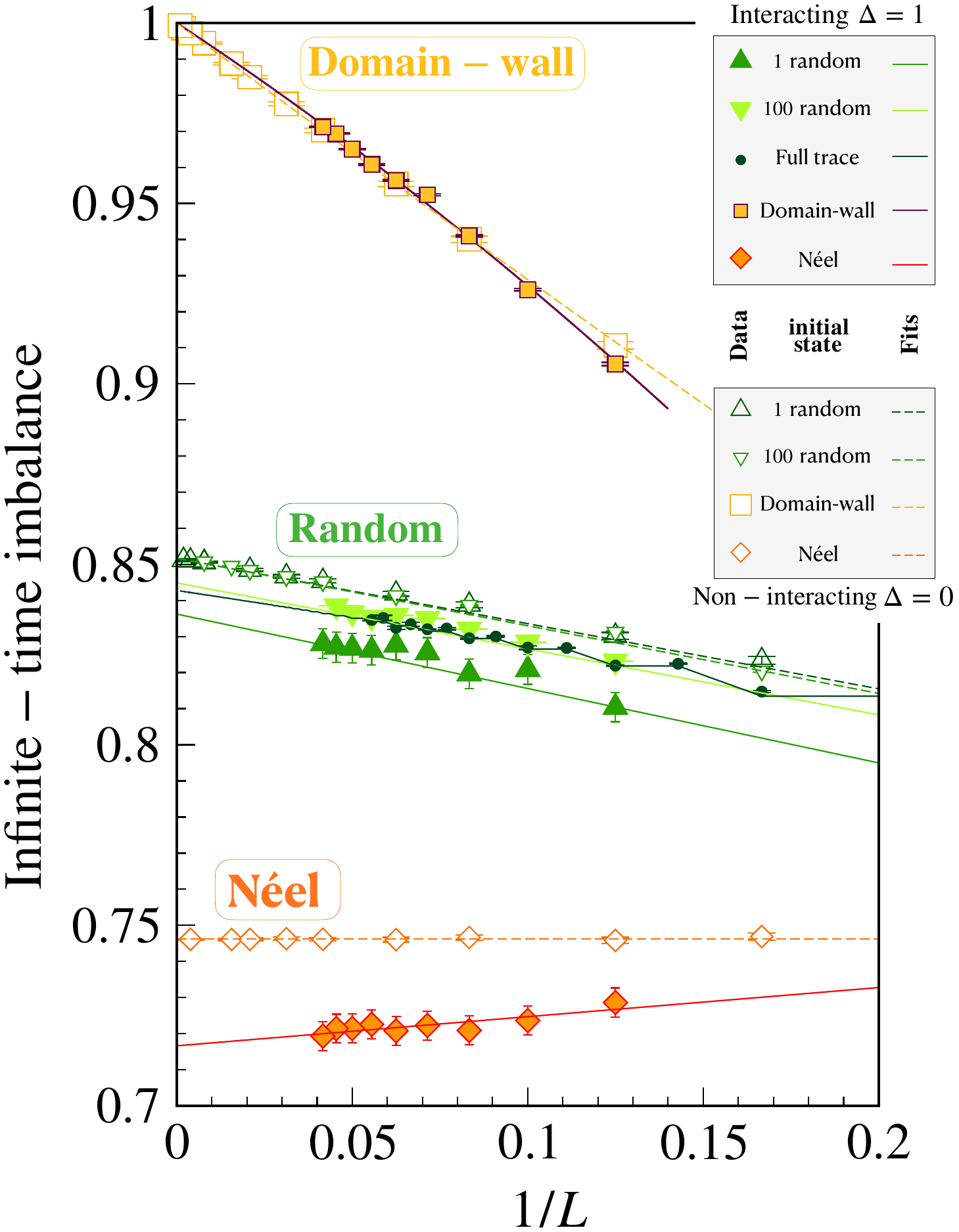}
    \caption{
Finite-size scaling of the long-time imbalance plotted as a function of $1/L$ for interacting $\Delta = 1$ (filled symbols) and non-interacting $\Delta = 0$ (open symbols) cases, for disorder $h=10$. Different initialization protocols are shown: green symbols correspond to averages over random initial states, obtained from either partial or full traces; yellow symbols denote the domain-wall initial state; and orange symbols correspond to the N\'eel initial state. The data illustrate the strong dependence of the long-time imbalance on both the interaction strength and the choice of initial conditions. Lines are fitting functions given in Tab.~\ref{tab:Ifits}.}
    \label{fig:FSS_Imb}
    \end{figure}

\paragraph{Random initial states --- } 
We first consider the dynamics from a single random initial state $\ket m$ for each disordered sample, taking the average afterwards. 
From the local-resonance toy model and for a pair of neighboring sites with similar fields ($\delta h \simeq 0$), we expect qualitatively different dynamics if $\ket m$ has corresponding polarized spins $|{\cal M}_{1-2}|=1$ ($\ket{\uparrow\uparrow} ; \ket{\downarrow\downarrow}$) or with $|{\cal M}_{1-2}|=0$   ($\ket{\downarrow\uparrow} ; \ket{\uparrow\downarrow}$) on this particular pair of sites, see Tab.~\ref{tab:Zinfty_2body}. Due to the global constraint $S_{\rm tot}^{z}=0$, it is slightly more probable to get the later case then the former. The probabilities of observing a particular orientation of the spin pair are simply obtained as:
\bea
p_{\uparrow\downarrow}=p_{\downarrow\uparrow}=\frac{1}{4}\left(1+\frac{1}{L-1}\right)\\
p_{\uparrow\uparrow}=p_{\downarrow\downarrow}=\frac{1}{4}\left(1-\frac{1}{L-1}\right).
\eea
Using the results of Tab.~\ref{tab:Zinfty_2body}, and taking the average over initial states and disorder simply yields
\bea
{\overline{Z_{\rm fluct.}(\infty)}}&=&\left(p_{\uparrow\uparrow}+p_{\downarrow\downarrow}\right)
+\left(p_{\uparrow\downarrow}+p_{\downarrow\uparrow}\right)\times \frac{1}{2}\times \frac{\Delta^2}{1+\Delta^2}\nonumber\\
&=&\frac{1}{2}+\frac{\Delta^2}{4(\Delta^2+1)}-\cfrac{\frac{1}{2}-\frac{\Delta^2}{4(\Delta^2+1)}}{L-1},
\label{eq:Zavg_toy}
\eea
which is exactly the same expression as if one does the full trace, see Eq.~\eqref{eq:2body_even}. Reinjecting this expression in the average imbalance Eq.~\eqref{eq:Iavg_toy}, we can rewrite it as
\be
{\cal I}_{\rm avg}^{\rm Rand}(L)\approx 1-x_\Delta\frac{\alpha}{h}
- x_{\Delta}\frac{\alpha}{h}
\frac{1}{L-1},\quad x_\Delta=\frac{\Delta^2+2}{4(\Delta^2+1)}.
\label{eq:Iavg_toy2}
\ee
This result is obviously also true if one performs the partial trace, averaging over a small finite set of initial states $N_s\ll N_{\rm H}$, as well as for the full trace ($N_s = N_{\rm H}$). 
We therefore conclude that the disorder-average imbalance, obtained either from taking the trace (full or partial) or starting from random initial states, is well described by Eq.~\eqref{eq:Iavg_toy2}, which is very well verified when comparing with the exact numerics shown in Fig.~\ref{fig:FSS_Imb}.

At this stage, a few comments are in order. First, we notice that Eq.~\eqref{eq:Iavg_toy2} remarkably captures the dependence of the average imbalance on the disorder strength $h$, the interaction $\Delta$, and the system size $L$. In the infinite size limit, the asymptotic ($L\to \infty$) average imbalance has corrections $\propto 1/h$ at strong disorder, in good agreement with Ref.~\cite{scoquart_scaling_2025}. A very interesting results concerns the finite size effects, which bring {\it negative} $1/L$ corrections, meaning that the thermodynamic value ${\cal I}_{\rm avg}(L=\infty)$ is reached from below, a result that remains true for any interaction strength $\Delta$. 
In the non-interacting limit, we find that the result obtained from free-fermion computations (Eq.~\eqref{eq:nonintrandom} in App.~\ref{sec:ff}) coincides exactly with Eq.~\eqref{eq:Iavg_toy}, upon identifying
\be
\frac{\alpha}{h}=2\exp(-1/\xi_{\rm eff}),
\label{eq:identification}
\ee
which yields the expected strong-disorder behavior for the localization length $1/\xi_{\rm eff} = \ln(h/h_0)$~\cite{colbois_breaking_2023}.

We clearly verify in Fig.~\ref{fig:FSS_Imb} the negative finite-size $1/L$ corrections in the non-interacting $\Delta=0$ case for several system sizes, as well at the  Heisenberg point $\Delta=1$ for $L=6,\,7,\,8,\,\ldots,\,17,\,18$. We also nicely observe a clear even-odd effect, in good agreement with the analytical prediction Eq.~\eqref{eq:2body_even} for the 2-body resonances.
The fits to a negative $1/L$ correction are summarized in Table~\ref{tab:Ifits} and allow to obtain excellent determinations of both the thermodynamic value of imbalance, and the strength of the $1/L$ corrections. Here we further emphasize that the three above protocols (full trace, partial average over $N_s=100$ initial states, or single random initial state) yield comparable  behaviors, in particular the striking result that the thermodynamic value ${\cal I}_{\rm avg}(L=\infty)$ should be reached from below.

We can go further by considering specific initial states that are not chosen randomly (e.g., domain-wall or N\'eel states), an approach commonly used in experiments and numerics.\\

\paragraph{Domain-wall initial state ---} Because of periodic boundary conditions, the less flippable basis state in the sector $S^z_{\rm tot}=0$ has necessarily two domain walls. We therefore consider the following initial state ${\ket{\rm 2DW}}={\ket{\uparrow\uparrow\uparrow\ldots\uparrow\downarrow\downarrow\downarrow\ldots\downarrow}}$. Building again on the statistical modeling given by Eq.~\eqref{eq:Iavg_toy}, we  evaluate the disorder average 
\be
{\overline{Z_{\rm fluct.}(\infty)}}=\frac{L-2}{L}+\frac{2}{L}\frac{\Delta^2}{\Delta^2+1},
\ee
where the first term in the rhs is the probability that a bond in a $L$-site chain has a polarized spin pair ($\uparrow\uparrow$ or $\downarrow\downarrow$) in the initial ${\ket{\rm 2DW}}$ state, while the second term accounts for anti-aligned spins, which occur with probability $2/L$. Injecting in Eq.~\eqref{eq:Iavg_toy}, we obtain
\be
{\cal{I}}_{\rm avg}^{\rm DW}\approx 1-\frac{2}{\Delta^2+1}\frac{\alpha}{h}\frac{1}{L}.
\label{eq:2IDW}
\ee
Interestingly, the disorder-average imbalance reaches its saturation value $=1$ in the thermodynamic limit, with negative finite-size corrections $\sim 1/L$, in perfect agreement with the numerics shown in Fig.~\ref{fig:FSS_Imb}. Eq.~\eqref{eq:2IDW} also describes the non-interacting case, which agrees with the free-fermion calculation Eq.~\eqref{eq:IDW} from App.~\ref{sec:ff}, at the order $1/h$, with the same identification given in Eq.~\eqref{eq:identification}.\\

\paragraph{N\'eel initial state ---} Starting from the most studied initial state, the N\'eel configuration $\ket{\uparrow \downarrow \uparrow \downarrow \ldots}$ (staggered charge-density wave in particle models), we can also use results from the two-site model of Sec.~\ref{sec:toy}. For this state, all two-site local configuration have ${\cal M}_{1-2}=0$ and furthermore are neighbored by spins with opposite direction. This means that only one contribution out of the four possible in Tab.~\ref{tab:Zinfty_2body} is present for the N\'eel state. Repeating the computation and assuming that averaging over disorder realizations provides the same combinatorial factors that from averaging over initial states, we expect
\be
{\overline{Z_{\rm fluct.}(\infty)}}\approx \frac{1}{4}\frac{\Delta^2}{\Delta^2+1}\left(1+\frac{1}{L}\right),
\ee
which gives an average infinite-time imbalance, starting from one of the two N\'eel configurations
\be
{\cal{I}}_{\rm avg}^{\rm N\acute eel}\approx 1-(1-y_\Delta)\frac{\alpha}{h}+y_\Delta\frac{\alpha}{h}\frac{1}{L},\quad y_\Delta=\cfrac{\Delta^2}{4(\Delta^2+1)}.
\label{eq:IavgNeel}
\ee
We first remark that the finite-size corrections $\sim 1/L$ have a different sign with respect to all the other situations discussed so far, meaning that in the special case of an initial N\'eel state, the thermodynamic limit imbalance is attained {\it{from above}} as $L$ increases. This finite-size effect is clearly observed in our numerics, see Fig.~\ref{fig:FSS_Imb} and the fitting form in Tab.~\ref{tab:Ifits}. In particular we also check that this finite-size effect is controlled by the interaction strength, $y_\Delta\sim \Delta^2$, which perfectly matches the absence of finite-size correction for free fermions (Eq.~\eqref{eq:INeel} in App.~\ref{sec:ff}), as seen in the numerical results up to $L=256$ sites. We also note that for $\Delta=0$, the strong disorder correction ${\cal{I}}_{\rm avg}^{\rm N\acute eel}(L=\infty)=1-{\alpha}/{h}$,
also perfectly matches the non-interacting calculation, once again using Eq.~\eqref{eq:identification}.

\begin{table}[h!]
  \centering
  \small
  \setlength{\tabcolsep}{2pt}          
  \renewcommand{\arraystretch}{1.8}    

  \adjustbox{max width=\linewidth}{%
    \begin{tabular}{lcc}
      \toprule
      Initial state & $\Delta = 0$ & $\Delta = 1$ \\
      \midrule

      N\'eel
        & $0.7461(2) + \frac{0.0001(26)}{L - 1}$
        & $0.717(7) + \frac{0.07(4)}{L - 1}$ \\

      Domain wall
        & $1 - \frac{0.73(2)}{L} + \frac{0.16(9)}{L^{2}}$
        & $1 - \frac{0.647(3)}{L} - \frac{0.81(5)}{L^{2}}$ \\

      \midrule

      \begin{tabular}[c]{@{}l@{}}
        Random (1 conf.) \\
        Random (100 conf.)
      \end{tabular}
        &
      \begin{tabular}[c]{@{}l@{}}
        $0.852(1) - \frac{0.14(1)}{L - 1}$ \\
        $0.8518(1) - \frac{0.152(2)}{L - 1}$
      \end{tabular}
        &
      \begin{tabular}[c]{@{}l@{}}
        $0.84(1) - \frac{0.16(9)}{L - 1}$ \\
        $0.8452(1) - \frac{0.149(1)}{L - 1}$
      \end{tabular}
        \\

      \midrule

      Full trace
        & \text{n.a.}
        & $0.843(2) - \frac{0.142(2)}{L - \left(1 + (-1)^L\right)/2}$ \\
      \bottomrule
    \end{tabular}%
  }

  \caption{Imbalance fitting functions and parameters shown in
  Fig.~\ref{fig:FSS_Imb} for the non-interacting ($\Delta = 0$) and
  interacting Heisenberg ($\Delta = 1$) cases, for disorder strength $h=10$.}
  \label{tab:Ifits}
\end{table}

\subsubsection{Discussion}

We now briefly discuss the consequences of these results (in particular Fig.~\ref{fig:FSS_Imb}) for quench dynamics in experiments and numerical simulations. Compared to a generic random initial state, the N\'eel configuration is known to be highly peculiar: all its bonds are of the form $\uparrow\downarrow$ or $\downarrow\uparrow$, and are therefore flippable under the XXZ Hamiltonian. In other words, in the spin-configuration space, the two N\'eel configurations present the largest possible connectivity, equal to $L$. This makes the N\'eel state maximally delocalizable and, as a consequence, not a generic initial state for probing the universal properties of the transition via quench dynamics. In a similar spirit, the domain-wall initial configuration is also non-generic, as it features only two flippable bonds (or even a single one in the case of open boundary conditions), i.e., a vanishing density, leading to Eq.~\eqref{eq:2IDW}.

Our analysis yields several estimates for the long-time average imbalance: Eq.~\eqref{eq:IavgNeel} for N\'eel, Eq.~\eqref{eq:2IDW} for domain-wall, and Eq.~\eqref{eq:Iavg_toy2} for generic random initial states. These results exhibit a strong dependence on the interaction strength $\Delta$, but perhaps even more surprising is the role played by finite-size corrections for the imbalance. Long recognized as a key ingredient in the MBL context, they provide here a new perspective on the interpretation of finite-size results. Indeed, only the dynamics starting from the N\'eel configuration display positive finite-size corrections, whereas generic random initial states lead to negative corrections, driven by few-body resonances. This behavior sharply contrasts with most finite-size results reported in MBL studies, as it represents a rare example of a quantity that appears more localized upon increasing the system size $L$.

\section{Conclusions}
In this work, we uncovered the existence of simple, local short-range resonance effects in the standard lattice model of many-body localization, which have direct consequences for the imbalance—the most widely used experimental measure of quantum localization.
 We find that, due to this local resonance effect, the different averages taken when computing or measuring the imbalance result in different marginal probability distributions. In particular, the probability distribution of the local autocorrelator ${\cal A}_j=\frac{1}{N_s}\sum_{m=1}^{N_s}  \langle Z_j(0)Z_j(t) \rangle $ in the limit of long-time displays secondary peak(s) which constitute direct signature(s) of this local resonance effect. This feature is captured by an analytically tractable few-site toy model. We have considered influence of finite time ($t$) and sampling ($N_s$) effects and verified that the main secondary peak can be  observed in realistic conditions (small values of $N_s$ and $t$), provided that the analysis of experimental data properly differentiates the different contributions when measuring average imbalance. This analysis is timely, as local measurement of individual occupation number is now readily available in experiments addressing many-body physics with ultracold atoms~\cite{Bakr_singleatom_Nature_2009,  
 Parsons_siteresolved_PRL_2015, Lukin_probing_science_2019,karch_probing_2025, hur_stability_2025,jongh_quantum_2025} and superconducting circuits~\cite{yao_observation_2023,li_many-body_2025, lunkin_evidence_2026}, which are platforms relevant to MBL.

This effect occurs with a probability that scales as $1/h$ ($h$ being the disorder strength) and its explanation in terms of a few-sites toy model requires the existence of well-polarized remainder of the sample (meaning, relatively strong MBL). In a disorder regime where these conditions are met, one important consequence is that short-range resonances govern the finite-size scaling of the average imbalance for the XXZ model Eq.~\eqref{eq:XXZ}. Perhaps quite counter-intuitively, the {\it infinite} (long)-time imbalance {\it increases} with system size (resulting from a negative $1/L$ correction) towards its thermodynamic value for generic random initial states. On the other hand, when considering quenches from the specific N\'eel state, the long-time imbalance {\it decreases} with system size (positive $1/L$ correction). This indicates that a certain degree of caution must be taken when interpreting numerical or experimental data of imbalance at {\it finite-}time. Indeed a slow decrease $\sim t^{-\beta}$ in a finite-time window finite-time of the imbalance of a finite-system, starting from a N\'eel state, does not necessarily imply long-time thermalization in the thermodynamic limit, but could instead simply signal a crossover {\it{from above}} to a smaller but finite asymptotic value ${\cal I}_{\rm avg}^{\rm N\acute eel}(L=\infty\,,t=\infty)\neq 0$. For instance, the value reported for the longest available time for the largest size in the  TEBD data of Ref.~\cite{sierant_challenges_2022} for disorder $h=10$ is ${\cal I}^{{\rm N\acute eel}}_{\rm avg} (t_{\rm max}\approx 10^3,\, L=200) \approx 0.7287$, which is quite larger than the expected one, ${\cal I}^{{\rm N\acute eel}}_{\rm avg}(t=\infty, L=200) \approx 0.717$ from Fig.~\ref{fig:FSS_Imb} and Tab.~\ref{tab:Ifits}. Therefore, a natural interpretation of the numerical results of Ref.~\cite{sierant_challenges_2022} is that the observed imbalance decay $\sim t^{-\beta}$ in the range $t\in [100,\,1000]$ could simply be a transient crossover towards a smaller asymptotic  value~\cite{Note3}

The strong influence of the initial state on behavior of the imbalance in the MBL phase was already pointed out (albeit for a different quasiperiodic model) in Ref.~\cite{prasad_initial_2022} where the dependence of the imbalance on the number of domain walls was linked to the initial state overlap with eigenstates of the Hamiltonian (participation ratio). Using a self-consistent Hartree-Fock approximate scheme, Ref.~\cite{weidinger_self_2018} also discussed the influence of the initial state but mostly on the ergodic side.
We also note the earlier work~\cite{hauschild_domain_2016}, which studied the evolution from a domain-wall initial state as a reliable probe of localization, connecting to 2D MBL experiments~\cite{Choi_2DMbl_2016}. Our study goes further by providing clear physical explanation and precise functional forms such as Eqs.~\eqref{eq:Iavg_toy2}, \eqref{eq:2IDW}, \eqref{eq:IavgNeel}.

We believe that the different finite-size effects on imbalance (and their dependence on the initial states) can also be probed experimentally and straightforwardly analyzed using these fitting forms. For instance, Ref.~\cite{wang_exploring_2025} reports the first experimental study of Stark many-body localization using a $12\times 2$-qubit superconducting processor, showing that the dynamics strongly depend on the initial state. Nevertheless, in very recent experimental studies of 2D MBL, the initial states were considered to be charge density waves~\cite{li_many-body_2025} or stripe patterns~\cite{hur_stability_2025}, similarly to the recent numerical study using tensor networks~\cite{humpert_tree_2025}. We believe that extending to other types of initial states would be very interesting in order to probe local resonances, and address some possible non-trivial finite-size scaling of the imbalance. In this respect, we note the very recent experimental work on 2D disordered systems~\cite{lunkin_evidence_2026}, in which random initial product states were employed.

In our study, the $1/L$ corrections can be traced back to the conservation of the total magnetization $S^z_{\rm tot}$, which arises from the absence of single spin-flip terms in the XXZ model considered here. It would be interesting to investigate local resonances and their impact on finite-size corrections to the imbalance in other models that host an MBL phase but do not conserve magnetization, such as the Imbrie model~\cite{Imbrie_2016, biroli_large-deviation_2024} and related non-symmetric models~\cite{scoquart_role_2024, scoquart_scaling_2025}, or the $\mathbb{Z}_2$-symmetric interacting Kitaev-Majorana chain~\cite{laflorencie_topological_2022}. Another natural extension of our work is to relax the assumption that spins near a resonance are perfectly frozen and to explore weaker disorder regimes where the main and secondary peaks of the long-time autocorrelator distribution begin to merge. Upon decreasing $h$, one could then study, across the MBL–ergodic transition, the evolution of the relaxation times of site-resolved autocorrelators.

\acknowledgments
We thank I. Bloch, M. Feigel'man, A. Mirlin, P. Sierant, M. Tarzia for very fruitful discussions. This work has been partly supported by the ANR research grant ManyBodyNet No. ANR-24- CE30- 5851, the EUR grant NanoX No. ANR-17-EURE0009 in the framework of the ”Programme des Investissements d’Avenir”, which is part of HQI initiative (www.hqi.fr) and is supported by France 2030 under the French National Research Agency award numbers ANR-22-PNCQ-0002 and ANR-23-PETQ-0002, and also benefited from the support of the Fondation Simone et Cino Del Duca. A. H. was supported by the Marie Sk\l{}odowska-Curie grant agreement No. 101110987. We acknowledge the use of HPC resources from CALMIP (grants 2025-P0677) and GENCI (projects A0150500225 and A0170500225), as well as of the PETSc~\cite{petsc-user-ref,petsc-efficient}, SLEPc~\cite{slepc-toms,slepc-users-manual}, MUMPS~\cite{MUMPS1,MUMPS2} sparse linear algebra libraries.

\vskip 1cm
\appendix
\setcounter{section}{0}
\setcounter{secnumdepth}{3}
\setcounter{figure}{0}
\setcounter{equation}{0}
\setcounter{table}{0}
\renewcommand\thesection{S\arabic{section}}
\renewcommand\thefigure{S\arabic{figure}}
\renewcommand\theequation{S\arabic{equation}}
\renewcommand\thetable{S\arabic{table}}

\onecolumngrid
\section{Details about time evolution}
\label{app:time}
\subsection{Krylov evolution}
\label{sec:Krylov}

Time evolution using Krylov space is an efficient numerical technique proposed by
Nauts and Wyatt~\cite{Nauts_Wyatt_Krylov} to approximate the action of the time-evolution operator $e^{-iHt} \ket{\psi_0}$, where $H$ is a Hermitian Hamiltonian acting on a high-dimensional Hilbert space and $\ket{\psi_0}$ is an initial state. Rather than computing the full matrix exponential, which is computationally infeasible for large many-body systems, the method constructs an $m$-dimensional Krylov subspace $\mathcal{K}_m(H, \ket{\psi_0}) = \text{span} \{ \ket{\psi_0}, H\ket{\psi_0}, H^2\ket{\psi_0}, \ldots, H^{m-1}\ket{\psi_0} \}$. 
Using the Lanczos algorithm for Hermitian $H$, an orthonormal basis $\{ \ket{v_j} \} $ of the Krylov subspace is built, and $H$ is projected onto this basis to yield a tridiagonal matrix  $ T_m $.
 The time-evolved state is then approximated as  
$ \ket{\psi(t)} \approx  V_m e^{-i T_m t} e_1$, 
where $ V_m \in \mathbb{C}^{N \times m}$
 is the matrix whose columns are the orthonormal basis vectors $\{ \ket{v_j} \} $
, and 
$e_1$
 is the first standard basis vector in ${R}^m$ and $ T_m = V_m^{ \dagger} H V_m $.
This projection reduces the exponential of a large operator to a small matrix exponential, which can be computed efficiently using standard techniques (e.g., diagonalization or Padé approximation). One can reach to the desired precision of convergence for the wavefunction by increasing the Krylov space dimension $m$. Krylov-based time evolution is particularly effective for short to intermediate times and can be integrated with adaptive schemes to ensure accuracy, making it highly suitable for quantum many-body dynamics. We have used the implementation provided in the SLEPc~\cite{slepc-toms,slepc-users-manual} package.

\subsection{Time averaged local magnetization}
\begin{figure}[h!]
    \centering
    \includegraphics[width=\columnwidth]{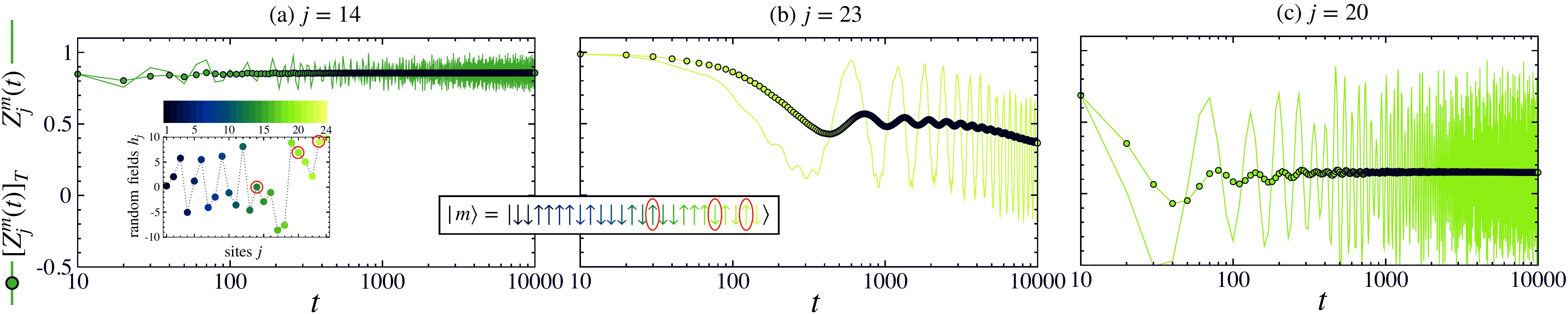}
\caption{Time evolution of the autocorrelator $Z_j^m(t)=\bra{m} \sigma^z_j (0) \sigma^z_j (t) \ket{m}$ together with its time-averaged form Eq.~\eqref{eq:Zmtime_app} with $t_{\rm min=10}$, shown for three representative cases from sample 1. The field distribution is shown in the inset, as well as the initial state $\ket m$, see also Fig.~\ref{fig:samples} in the main text. (a) Typical localized response; (b) slowly decaying magnetization; and (c) fast relaxation to zero.}
    \label{fig:local}
\end{figure}

In order to smooth the curves and get rid of fast oscillations, we performed a time-average of the autocorrelator data 
\be
[Z_j^m(t)]_{T}=\frac{1}{t-t_{\rm min}}\int_{t_{\rm min}}^{t}Z_j^m(\tau){\rm{d}}\tau.
\label{eq:Zmtime_app}
\ee
Using the discrete version of the above equation, we exemplify the procedure in Fig.~\ref{fig:local}, which shows three representative behaviors that appear in sample 1 (Fig.~\ref{fig:samples} (a) of the main text): (a) Typical localized behavior ;  (b) Slow decay that may eventually loose completely its initial state polarization at very long time ;  (c) Rapid decay to zero. In this last case, which is representative of a short-range resonance, the magnetization is typically described by a Rabi-like behavior of the form Eq.~\eqref{eq:Zmt_app}, and the time averaging efficiently allows to describe the long-time expectation Eq.~\eqref{eq:Ztavg} with ${\cal G}\approx 0$.

\section{Toy models for short-range resonances: analytical details}
\label{sec:AppA}
\subsection{2-site resonances}
We first 
consider a toy model that consists of 2 sites embedded in a trivially localized product-state background. We describe this by the following "toy-state" ansatz 
\be
{\ket{\rm TS}}={\ket{\uparrow\uparrow\downarrow\ldots}}\otimes{\ket{\circ_1~\circ_2}}\otimes {\ket{\downarrow\uparrow\downarrow\ldots}},
\label{eq:TS_app}
\ee
where ${\ket{\circ_1~\circ_2}}$ is the pair of (potentially fluctuating) sites for which we want to characterize the quantum dynamics. Assuming that the neighboring sites of the pair are frozen and do not fluctuate, we can write a 2-site Hamiltonian for sites 1 and 2, which  takes the simple form
\be
{\cal{H}}_{\rm eff}=
\sigma^x_1\sigma^x_{2}+\sigma^y_1\sigma^y_{2}+\Delta \sigma^z_1\sigma^z_{2}-2\sum_{j=1}^{2}h_{j}^{\rm eff} \sigma^z_j.
\label{eq:Heff2_app}
\ee
Here the effective fields $h_{1,2}^{\rm eff}=h_{1,2}\pm \frac{\Delta}{2}$ 
account for the effects of external random fields and nearest-neighbor interactions, the sign depending on the neighboring spin configurations. It is straightforward to diagonalize this two-body problem in the 4-dimensional Hilbert space spanned by the 4 basis states 
$\{{\ket{\uparrow\uparrow}};{\ket{\downarrow\downarrow}}; {\ket{\uparrow\downarrow}};{\ket{\downarrow\uparrow}}\}$.

\subsubsection{Polarized states} For the finite magnetization sectors $S_1^z+S_2^z=\pm 1$ each of dimension $1$, ${\ket{\uparrow\uparrow}}$ and ${\ket{\downarrow\downarrow}}$ are the trivial eigenstates. If an initial state $m_0$ has its local projection on sites 1 and 2 that belongs to this subset, one expects the local magnetizations to remain constant over time, i.e.  $Z^{m_0}_{1,2}(t)\approx 1$, $\forall t$.\\

\subsubsection{Zero magnetization sector} The states 
${\ket{\uparrow\downarrow}}$ and ${\ket{\downarrow\uparrow}}$ form a 2-dimensional subset with $S_1^z+S_2^z=0$. Diagonalizing ${\cal{H}}_{\rm eff}$ boils down to solving the $2\times 2$ matrix

\bea
   M=
  \left[ {\begin{array}{cc}
   -{\cal{G}} & 1 \\
   1 & {\cal{G}} \\
  \end{array} } \right].
\eea
${\cal{G}}$ is the gap between ${\ket{\uparrow\downarrow}}$ and ${\ket{\downarrow\uparrow}}$, see Tab.~\ref{tab:1_app}.
\begin{table}[b!]
\begin{center}
\begin{tabular}{l|c|c|c|c|}
States \textbackslash ~ NN. Config. & $\uparrow~.~.~\uparrow$ & $\downarrow~.~.~\downarrow$ & $\uparrow~.~.~\downarrow$ & $\downarrow~.~.~\uparrow$\\
\hline
${\ket{\uparrow\uparrow}}$ & $\frac{3J\Delta}{4}+\frac{(h_1+h_2)}{2}$ & $-\frac{J\Delta}{4}+\frac{(h_1+h_2)}{2}$ & $\frac{J\Delta}{4}+\frac{(h_1+h_2)}{2}$ & $\frac{J\Delta}{4}+\frac{(h_1+h_2)}{2}$\\
${\ket{\downarrow\downarrow}}$ & $-\frac{J\Delta}{4}-\frac{(h_1+h_2)}{2}$ & $\frac{3J\Delta}{4}-\frac{(h_1+h_2)}{2}$ & $\frac{J\Delta}{4}-\frac{(h_1+h_2)}{2}$ & $\frac{J\Delta}{4}-\frac{(h_1+h_2)}{2}$\\
${\ket{\uparrow\downarrow}}$ & $-\frac{J\Delta}{4}+\frac{(h_1-h_2)}{2}$ & $-\frac{J\Delta}{4}+\frac{(h_1-h_2)}{2}$ & $\frac{J\Delta}{4}+\frac{(h_1-h_2)}{2}$ & $-\frac{3J\Delta}{4}+\frac{(h_1-h_2)}{2}$\\
${\ket{\downarrow\uparrow}}$ & $-\frac{J\Delta}{4}-\frac{(h_1-h_2)}{2}$ & $-\frac{J\Delta}{4}-\frac{(h_1-h_2)}{2}$ & $-\frac{3J\Delta}{4}-\frac{(h_1-h_2)}{2}$ & $\frac{J\Delta}{4}-\frac{(h_1-h_2)}{2}$\\
\hline
\end{tabular}
\caption{Energies of the the 16 possible configurations for the 2-site toy model whose 4 states are in the left column and the 4 possible nearest neighbor configurations are indicated in the top line.}
\label{tab:1_app}
\end{center}
\end{table}
There are 4 different configurations of neighboring spins, which yields ${\cal{G}}=\delta h,\,\delta h,\,\Delta\pm\delta h$, where $\delta h=|h_{1}^{\rm eff}-h_{2}^{\rm eff}|$. The eigenenergies $\pm \Omega$ are given by $\Omega = \sqrt{1+{\cal{G}}^2}$, and the eigenstates can be written as follows
\begin{equation}
\label{eq:psi}
\begin{array}{l}
{\ket{\psi_{+}}}=\cos({\theta}/{2}){\ket{\uparrow\downarrow}}+ \sin({\theta}/{2}){\ket{\downarrow\uparrow}}\\
{\ket{\psi_{-}}}=\sin({\theta}/{2}){\ket{\uparrow\downarrow}}- \cos({\theta}/{2}){\ket{\downarrow\uparrow}},
\end{array}
\end{equation}
where $\cos \theta = {\cal G}/{\sqrt{1+{\cal{G}}^2}}$.
The time evolution of the local magnetizations, starting for instance from ${\ket{\uparrow\downarrow}}$,
\be
\langle \sigma_{1,2}^z(t)\rangle={\bra{\uparrow\downarrow}}{\rm{e}}^{{\rm i}{\cal{H}}_{\rm eff}t} \sigma_{1,2}^z {\rm{e}}^{-{\rm i}{\cal{H}}_{\rm eff}t}{\ket{\uparrow\downarrow}},
\ee
is easy to express using the decomposition of the initial state in the eigenbasis ${\ket{\uparrow\downarrow}}=\cos(\frac{\theta}{2}){\ket{\psi_+}}+\sin(\frac{\theta}{2}){\ket{\psi_-}}$. We therefore get the very simple expression
\be
Z_{1,2}^m(t)=1-\frac{2}{1+{\cal{G}}^2}\sin^2(\Omega t),
\label{eq:Zmt_app}
\ee 
which yield the time-averaged autocorrelator at long time 
\be 
\left[Z_{1,2}^m(t\gg 1)\right]_{T} \longrightarrow \frac{{\cal{G}}^2}{1+{\cal{G}}^2}.
\label{eq:Ztavg_app}
\ee

The final results for the the time-averaged autocorrelator in the long-time limit $[Z_{1,2}^{m}(t\gg 1)]_T$ for the two sites $j=1,2$ of the toy model, which depend on the four possible initial configurations for the sites $j=1,\,2$ and the four different nearest-neighbor spin states, are summarized in Tab.~\ref{tab:Zinfty_2body_tab} 
\begin{table}[h!]
  \centering
  \setlength{\tabcolsep}{8pt}    
  \renewcommand{\arraystretch}{1.2} 
  \begin{tabular}{lcc}
    \toprule
    & \multicolumn{2}{c}{Configurations of the neighbours} \\
    \cmidrule(lr){2-3}
    Initial states 
      & $(\uparrow \cdots \uparrow)$ or $(\downarrow \cdots \downarrow)$
      & $(\uparrow \cdots \downarrow)$ or $(\downarrow \cdots \uparrow)$ \\
    \midrule
    $\ket{\uparrow\downarrow}$ or $\ket{\downarrow\uparrow}$
      & $\dfrac{\delta h^2}{1 + \delta h^2}$
      & $\dfrac{(\Delta \pm \delta h)^2}{1 + (\Delta \pm \delta h)^2}$ \\
    $\ket{\uparrow\uparrow}$ or $\ket{\downarrow\downarrow}$
      & $1$ & $1$ \\
    \bottomrule
  \end{tabular}

\caption{Same as the first Tab. of the main text. Toy model results for the asymptotic values $[Z_{j}^{m}(t\gg 1)]_T$ of the time-averaged autocorrelator in the long-time limit for the two sites $j=1,2$ of the toy model Eq.~\eqref{eq:TS_app} and Eq.~\eqref{eq:Heff2_app}. The 16 possible spin configurations, involving the 2 central sites (left column) and the 2 neighbors (top line), are taken into account.}
\label{tab:Zinfty_2body_tab}
\end{table}
\subsection{Numerical checks of 2-body resonances}

\begin{figure}[t!]
    \centering
    \includegraphics[width=.45\columnwidth]{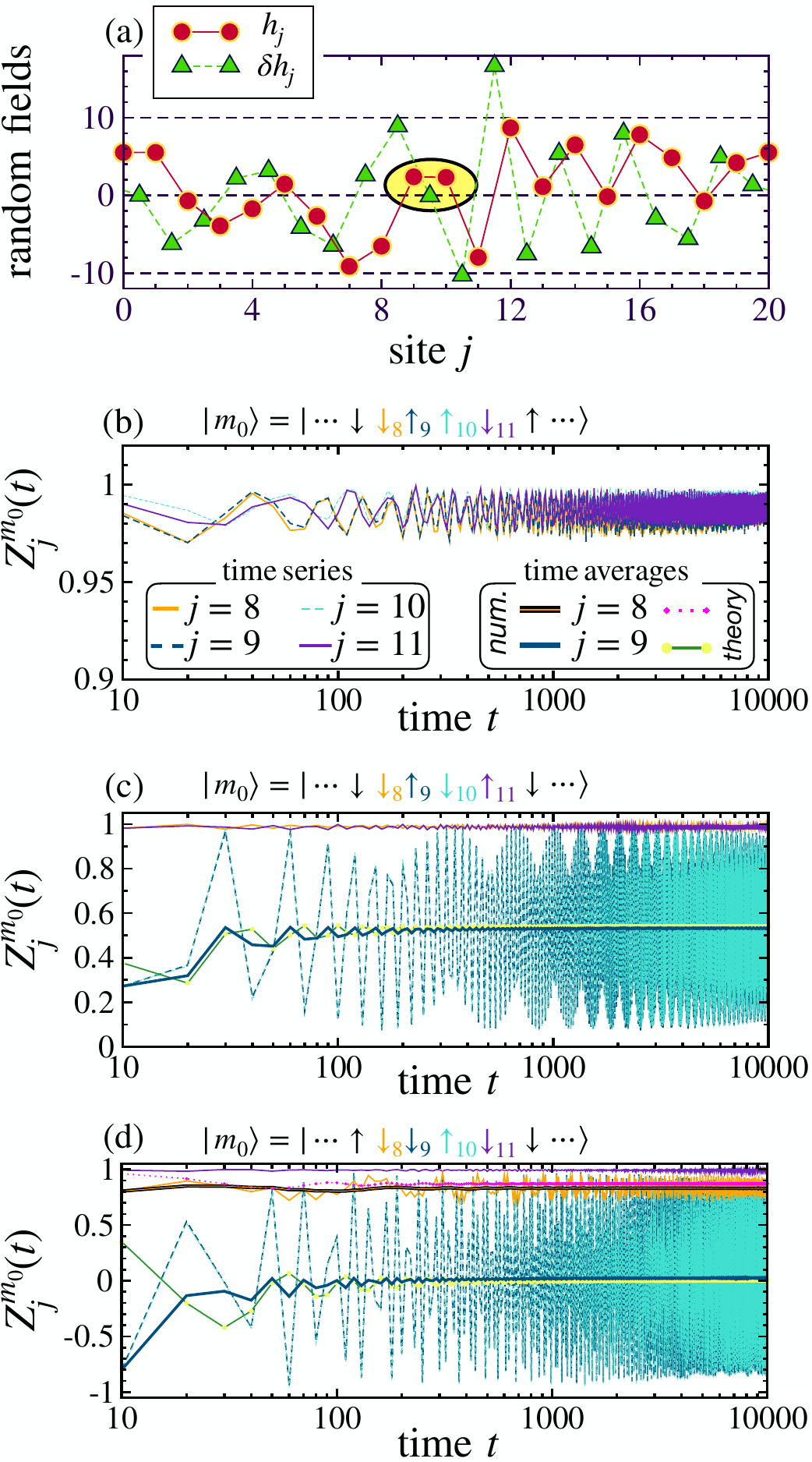}
    \caption{Numerical validation of the toy-model description of two-body resonances using a typical disordered sample ($h=10$). Panel (a) shows the random field configurations $\{h_j\}$ of this $L=20$ sample, together with the bond variables $\delta h_j = h_{j+1} - h_j$. The focus is made on the resonating sites $j=9-10$, together with their left and right neighbors $j=8,\,11$. Panels (b-d) shows time series for these 4 sites (thin dashed and full lines) for 3 specific initial (product) states ${\ket{m_0}}$ indicated on top of each panel. Time averages are also shown in panels (c-d) for $j=8-9$ (thick lines). The theoretical predictions Eq.~\eqref{eq:Zmt_app}, after  numerical integration of Eq.~\eqref{eq:Ztavg_app}, are also shown. The comparison is excellent.}   
    \label{fig:L20-h10_seed265_2body}
\end{figure}

Two-body resonances can be numerically observed using another $L=20$ sample, whose random field configuration is shown in Fig.~\ref{fig:L20-h10_seed265_2body}(a), along with the bond variable $\delta h_j = h_{j+1} - h_j$, which represents the field difference between nearest neighbors. We focus on sites $j = 9$ and $10$, which exhibit a very small bond variable $|\delta h| \approx 0.08302$, whereas the neighboring bonds are much stiffer, with $|\delta h| \sim 10$. Figs.~\ref{fig:L20-h10_seed265_2body}(b–d) display the autocorrelators $Z_j^m(t)$ for four specific sites $j = 8,\,9,\,10,\,11$, using 3 representative initial states $\ket{m_0}$ (indicated on the plots), which illustrate the main different cases displayed in Tab.~\ref{tab:Zinfty_2body_tab}.

\subsubsection{Fully polarized states} The polarized case is shown in panel (b) for an initial state ${\ket{~\cdots~\uparrow_9\uparrow_{10}~\cdots~}}$. As expected, the polarizations remain very close to full saturation at all time, which validates the picture of the finite magnetization subsector. 

\subsubsection{Fluctuating sites} In panel (c) and (d) the local projection of the initial state on sites $9-10$ belongs to the zero magnetization subspace, for which the local magnetization dynamics strongly depends on the local gap ${\cal G}=\delta h$ or $\Delta\pm\delta h$ that is determined by the neighboring configurations. The field difference $\delta h$ of the considered pair $9-10$ being very small, we anticipate the local magnetization dynamics to strongly fluctuate, as expected from Eq.~\eqref{eq:Zmt_app}. The time-averaged magnetizations are found to be in excellent agreement with the analytical prediction Eq.~\eqref{eq:Ztavg_app}, see also Tab.~\ref{tab:Zinfty_2body_tab}. In panel (d) we additionally analyze the dynamics of the neighboring site $j=8$, which shows a long-time magnetization saturating well below one, at $\approx 0.83$. Applying the toy model description to the pair of sites $7-8$, which for the particular initial state has a local gap ${\cal{G}}\simeq 2.6$, we expect a long-time asymptotic value from Eq.~\eqref{eq:Ztavg} $\approx 0.87$, which is in reasonable agreement (below $5\%$ error) with the exact simulation result.

\subsection{Three-body resonances}
\label{sec:3br}
One can use a similar approach for 3 consecutive sites in a locally uniform region. The analytical calculation is still possible but a bit long, so we will only quote the main results here. In order to deal with simple expressions, we assume that local random fields are equal ($h_1= h_2= h_3= h$). The 3-site effective Hamiltonian is
\be
{\cal{H}}_{\rm eff}^{(3)}=\sum_{j=1}^{2}
\sigma^x_j\sigma^x_{j+1}+\sigma^y_j\sigma^y_{j+1}+\Delta \sigma^z_j\sigma^z_{j+1}
-2\sum_{j=1}^{3}h_{j}^{\rm eff} \sigma^z_j,
\label{eq:Heff}
\ee
where effective magnetic fields are simply given by $h_2^{\rm eff}=h$, and $h_{1,3}^{\rm eff}=h+{\cal{B}}_{\Delta}$, where the term ${\cal{B}}_{\Delta}$ 
accounts for the effects of nearest-neighbor interactions at the boundaries of the 3-site clusters. This mean-field treatment of the interaction at both edges is justified by our assumption that, in the presence of strong disorder, the fluctuating clusters that we aim to describe are surrounded by strongly pinned moments that are expected to fluctuate very weakly.

The middle $j=2$ and boundary $j=1,\,3$ sites will display distinct behaviors, and after a (lengthy but) straightforward calculation, we arrive at the following expressions for the infinite-time limit of the (fully traced) autocorrelators ${\cal{A}}_{j}^{\rm Full}(\infty)$, here given for even $L$ (extension to odd chains is straightforward):
\be
{\cal{A}}_{\rm 3br,\, mid}^{\rm Full}(\infty)\approx \frac{1}{2}+\Delta^2 f_{m}(\Delta)-\frac{1/2-\Delta^2f_{m}(\Delta)}{L-1},
\ee
with 

\be
f_{m}(\Delta)=1/(8\Delta^2+16)-7/(64\Delta^4+52\Delta^2+64),
\ee
for middle sites. For the two boundary sites we obtain a different result
\be
{\cal{A}}_{\rm 3br,\, left/right}^{\rm Full}(\infty)\approx \frac{3}{8}+\Delta^2 f_b(\Delta)
-\frac{5/8-\Delta^2 f_b(\Delta)}{L-1}
\ee
with 
\be
f_b(\Delta)={1}/({32\Delta^2+64})+({48\Delta^2+9})/({256\Delta^4+208\Delta^2+128}).
\ee
\section{Non-interacting case: Free-fermion computations}
\label{sec:ff}

Here we specialize to the non-interacting limit $\Delta=0$ of the XXZ Hamiltonian Eq.~\eqref{eq:XXZ}. Using the Jordan--Wigner transformation, the model maps to a system of free fermions,
\begin{align}
    \mathcal{H} = \sum_{i=1}^L h_i c_i^\dagger c_i + \frac{1}{2} \sum_{i=1}^L \left( c_i^\dagger c_{i+1} + h.c.\right),
\end{align}
for which all single-particle eigenstates exponentially localized. The many-body eigenstates are formed by Slater determinants of these single particle orbitals. In this non-interacting limit, disorder-averaged fermionic fields are Gaussian, and owing to Wick's theorem, the quench dynamics of the system starting from any product state
$\ket{m}$ is fully encoded in the one-body density matrix
\begin{align}
C_{ij}(t)=\bra{\Psi_m(t)}c_i^\dagger c_j\ket{\Psi_m(t)},\qquad 
\ket{\Psi_m(t)}=e^{-i\mathcal{H}t}\ket{m}. 
\end{align}
The long time-average of $C_{ij}(t)$ is obtained from the diagonal ensemble: writing the Hamiltonian in its single-particle eigenmode basis
$\mathcal{H}=\sum_{n=1}^L \varepsilon_n\,\alpha_n^\dagger\alpha_n$ with
$\alpha_n=\sum_{i=1}^L \psi_n(i)c_i$, the one-body density matrix at infinite time obeys
\begin{equation}
C_{ij}(\infty)  
= \delta_{ij}\bra{\Psi_m(\infty)}c_i^\dagger c_j\ket{\Psi_m(\infty)}
= \delta_{ij}\sum_{n=1}^L |\psi_n(j)|^2 \sum_{k=1}^L |\psi_n(k)|^2\, n_k^m ,
\end{equation}
where $n_k^m=\bra m c_k^\dagger c_k\ket m=(1+\zeta_k^m)/2$, with $\zeta_j^m=\bra m\sigma_j^z\ket m=\pm1$ the local magnetization of the initial product state. The local diagonal autocorrelator~\eqref{eq:Zm} introduced in the main text admits the expression
\begin{align}
Z_j^m(t) = \bra{m}\sigma_j^z(0)\,\sigma_j^z(t)\ket{m}
         = \zeta_j^m\,\bra{\Psi_m(t)}\sigma_j^z\ket{\Psi_m(t)}.
\label{eq:Zm_noninteracting}
\end{align}
Using $\sigma_j^z = 2 c_j^\dagger c_j - 1$ and the completeness of the single-particle eigenfunctions, we find
\begin{equation}
\bra{\Psi_m(\infty)}\sigma_j^z\ket{\Psi_m(\infty)}
= \sum_{n=1}^L |\psi_n(j)|^2 \sum_{k=1}^L |\psi_n(k)|^2\,\zeta_k^m ,
\end{equation}
and therefore
\begin{equation}
Z_j^m(\infty)
= \zeta_j^m \sum_{n=1}^L |\psi_n(j)|^2 
                   \sum_{k=1}^L |\psi_n(k)|^2\,\zeta_k^m .
\label{eq:Zinf_correct}
\end{equation}
We now connect Eq.~\eqref{eq:Zinf_correct} to the site-resolved autocorrelators ${\cal A}_j(t)$ defined in Sec.~\ref{sec:state-averaging-example} of the main text.
Recall that
\begin{equation}
{\cal{A}}^{\rm Full}_j(t)=\frac{1}{N_{\rm H}}\sum_{m=1}^{N_{\rm H}}Z^{m}_j(t),
\label{eq:Aj_full_def}
\end{equation}
corresponds to a full trace, or ``infinite temperature'' average over all product states. 
Inserting Eq.~\eqref{eq:Zinf_correct} into Eq.~\eqref{eq:Aj} gives, at infinite time,
\begin{equation}
{\cal A}_j^{\rm Full}(\infty)
= \sum_{n=1}^L |\psi_n(j)|^2 \sum_{k=1}^L |\psi_n(k)|^2\,
C_s^{\rm Full}(j,k). 
\label{eq:Aj_partial_from_Z}
\end{equation}
The initial-state dependence enters only through the averaged spin correlator that is defined as
\begin{equation}
C_s^{\rm Full}(j,k)
= \frac{1}{{ N}_H}\sum_{m=1}^{{ N}_H} \zeta_j^m\zeta_k^m,
\label{eq:Cs_subset}
\end{equation}
where the sum runs over the set of all product states in the zero-magnetization sector. By translation invariance on the ring (we consider periodic boundary conditions), the averaged spin correlator depends only on the separation $r=k-j$:
\begin{equation}
C_s^{\rm Full}(j,k) \equiv C_s^{\rm Full}(r),
\end{equation}
with $r=(k-j)\,{\rm mod}\,L$. A straightforward combinatorial calculation for uniformly random half-filled configurations yields, in the limit $L\gg 1$:
\begin{equation}
C_s^{\rm Full}(r)=
\begin{cases}
1, & r=0,\\[3pt]
-\dfrac{1}{L-1}, & r\neq 0,
\end{cases}
\label{eq:Cs_full_half_filled}
\end{equation}
reflecting the fact that the magnetization constraint $\sum_j \zeta_j^m=0$ enforces
$\sum_{r=1}^{L-1} C_s^{\rm Full}(r)=-1$. Using Eq.~\eqref{eq:Cs_full_half_filled} in Eq.~\eqref{eq:Aj_partial_from_Z}, one finds
\begin{align}
{\cal A}_j^{\rm Full}(\infty)
&= \sum_{n=1}^L |\psi_n(j)|^2 
\left[
|\psi_n(j)|^2
+ \sum_{k\neq j} |\psi_n(k)|^2 \left(-\frac{1}{L-1}\right)
\right]\nonumber\\
&= \sum_{n=1}^L |\psi_n(j)|^2 
\left[
|\psi_n(j)|^2
- \frac{1}{L-1}\left(1-|\psi_n(j)|^2\right)
\right].
\label{eq:Aj_full_modeform}
\end{align}
While Eq.~\eqref{eq:Aj_full_modeform} is exact for a fixed disorder realization, it becomes more transparent after averaging over disorder. In this case, translational invariance implies that ${\cal A}_j^{\rm Full}(\infty)$ is independent of $j$, and it is convenient to re-express it in terms of the localization kernel introduced below.\\

\noindent\textbf{Disorder averaging.} Let us now consider the disorder-averaged localization kernel
\begin{align}
g(r) \equiv \frac{1}{L}\sum_{j=1}^L \sum_{n=1}^L 
\overline{|\psi_n(j)|^2\,|\psi_n(j+r)|^2}, 
\label{eq:kernel_def_appendix}
\end{align}
 which measures the typical spatial overlap, at separation $r$, of the intensity profile of a single eigenstate. In a localized regime, single-particle eigenstates are concentrated
around a localization center and their intensities decay on the scale of the
(energy-dependent) localization length $\xi_n$. As a result, the probability for a given eigenstate to carry significant weight simultaneously at two sites separated by $r$ is suppressed exponentially. For a fixed eigenstate $n$ (energy $E_n$),
this suggests the scaling
\begin{equation}
\frac{1}{L}\sum_{j=1}^L \overline{|\psi_n(j)|^2|\psi_n(j+r)|^2}
\sim A_n\,e^{-2r/\xi(E_n)},
\end{equation}
up to short-distance corrections and subexponential prefactors. Summing over eigenstates then yields a weighted mixture of exponentials, which we parametrize by an effective length $\xi_{\rm eff}$ appropriate to the chosen spectral window and distance range:
\begin{equation}
g(r)\sim e^{-2r/\xi_{\rm eff}}.
\end{equation}
On a lattice, it is convenient to parametrize the exponential decay by the form $g(r)\simeq A q^r$, where $q=e^{-2/\xi_{\rm eff}}$, and the prefactor $A$ absorbs non-universal short-distance details. In our case, we normalize $g$ as
\begin{equation}
g(r)\simeq (1-q)q^r,
\label{eq:kernel_form_appendix}
\end{equation}
such that it satisfies the sum rule $\sum_{r=0}^{L-1} g(r)= 1$.
Using Eq.~\eqref{eq:Zinf_correct}, summing over sites $j$, and averaging over disorder, one finds the general relation
\begin{equation}
{\cal I}_{\rm avg}(\infty) = \frac{1}{L}\sum_{j=1}^L Z_j^m(\infty)
= \sum_{r=0}^{L-1} g(r)\,C_s^m(r),
\label{eq:I_from_Cs_appendix}
\end{equation}
with $C_s^m(r)=\frac{1}{L}\sum_{j=1}^L \zeta_j^m\zeta_{j+r}^m$ the spatial autocorrelator of a given initial configuration $m$. In particular, for the full trace over random half-filled product states, averaging Eq.~\eqref{eq:I_from_Cs_appendix} over initial states replaces $C_s^m(r)$ by $C_s^{\rm Full}(r)$ given in Eq.~\eqref{eq:Cs_full_half_filled}.  
Since ${\cal A}_j^{\rm Full}(\infty)$ is independent of $j$ after disorder averaging, Eq.~\eqref{eq:Aj_full_def} implies that, upon taking disorder average,
\begin{equation}
{\cal A}_j^{\rm Full}(\infty) = {\cal I}_{\rm avg}(\infty)
= \sum_{r=0}^{L-1} g(r)\,C_s^{\rm Full}(r).
\end{equation}
Using Eqs.~\eqref{eq:Cs_full_half_filled} and \eqref{eq:kernel_form_appendix}, we obtain
\begin{align}
{\cal A}_j^{\rm Full}(\infty)
&=g(0) + \sum_{r=1}^{L-1} g(r)\left(-\frac{1}{L-1}\right)
\nonumber\\
&= g(0) - \frac{1-g(0)}{L-1}\nonumber\\
&= (1-q) - \frac{q}{L-1},
\label{eq:nonintrandom}
\end{align}
which coincides with the system size scaling of the infinite-time averaged imbalance obtained from averaging the dynamics over random half-filled initial states (see the discussion around Fig.~\ref{fig:FSS_Imb} in the main text).\\

\noindent\textbf{Néel initial state.}
For the initial Néel antiferromagnetic state, which corresponds to alternating fermion occupancies of $0$ and $1$, one has $\zeta_j^m=(-1)^j$, such that $C_s^{\rm Néel}(r)=(-1)^r$ for all $r$.
Inserting into Eq.~\eqref{eq:I_from_Cs_appendix} gives
\begin{equation}
\mathcal{I}^{\rm N\acute eel}(\infty)
=(1-q)\sum_{r=0}^\infty (-q)^r=\frac{1-q}{1+q}
= \tanh(1/\xi_{\rm eff}).
\label{eq:INeel}
\end{equation}
This result reproduces the constant behavior of the infinite-time imbalance following a Néel quench, observed numerically in Fig.~\ref{fig:FSS_Imb}. Note that the saturation value depends only on the effective localization length $\xi_{\rm eff}$.\\

\noindent\textbf{Single domain-wall initial state.}
Consider a initial state configuration with a single domain wall, that is $L/2$ consecutive sites with occupancy $0$ followed by $N/2$ occupied sites. Then
$\zeta_j^m=+1$ for $j\in[1,L/2]$ and $-1$ otherwise. One finds
\begin{equation}
C_s^{\rm DW}(r) = 1 - \frac{4r}{L},
\label{eq:CDW}
\end{equation}
which is valid for $1\ll r\ll L$. Substituting into Eq.~\eqref{eq:I_from_Cs_appendix}, one obtains
\begin{equation}
\mathcal{I}^{\rm DW}(\infty)
\simeq (1-q) \left(1- \frac{4}{L}\frac{q}{1-q}\right)= 1 - \frac{4}{L}\frac{e^{-2/\xi_{\rm eff}}}{1-e^{-2/\xi_{\rm eff}}}.
\label{eq:IDW}
\end{equation}
In contrast to the Néel state, the imbalance has a {\it negative} finite size correction decaying with $1/L$, also in agreement with the numerics presented in Fig.~\ref{fig:FSS_Imb}.
\twocolumngrid
\bibliography{biblio_new}
\end{document}